\documentclass[numberedappendix,appendixfloats,iop,apj]{emulateapj}

\usepackage[usenames,dvipsnames]{xcolor}
\usepackage{times, apjfonts}
\usepackage{amsmath, amsthm, amssymb}
\usepackage{graphicx, epsfig}
\usepackage{natbib}
\usepackage{verbatim}
\usepackage{epstopdf} 
\usepackage{CJK} 
\usepackage[colorlinks=true, anchorcolor=Sepia, linkcolor=red, urlcolor=OliveGreen, citecolor=blue, breaklinks=true]{hyperref}
\usepackage[hyphenbreaks]{breakurl}
\usepackage{datetime}
\usepackage[all]{hypcap}
\def\sectionautorefname~#1\null{Section #1\null}
\def\subsectionautorefname~#1\null{Section #1\null}
\def\subsubsectionautorefname~#1\null{Section #1\null}

\slugcomment{Accepted for publication in ApJ}
\shorttitle{\ion{C}{4}-based SE BH mass estimators}
\shortauthors{Park et al.}

\newcommand{\mbh}{\ensuremath{M_{\rm BH}}}

\newcommand{\msigma}{$M_{\rm BH}-\sigma_{*}$}

\newcommand{\msun}{\ensuremath{M_{\odot}}}
\newcommand{\kms}{km~s$^{\rm -1}$}
\newcommand{\ergs}{erg~s$^{\rm -1}$}

\newcommand{\linedisp}{$\sigma_{\rm line}$}
\newcommand{\iron}{\ion{Fe}{2}}
\newcommand{\FeII}{\ion{Fe}{2}}
\newcommand{\Hb}{H$\beta$}
\newcommand{\Ha}{H$\alpha$}
\newcommand{\CIV}{\ion{C}{4}}
\newcommand{\MgII}{\ion{Mg}{2}}
\newcommand{\HST}{{\it HST}}
\newcommand{\IUE}{{\it IUE}}

\begin{document}
\begin{CJK*}{UTF8}{mj}

\title{Extending the Calibration of \CIV-Based Single-Epoch Black Hole
  Mass Estimators for Active Galactic Nuclei$^{\star}$}

\author{Daeseong Park (박 대 성 )$^{1,2,3,\dag}$}
\author{Aaron J. Barth $^{3}$}
\author{Jong-Hak Woo (우 종 학 )$^{4}$}
\author{Matthew A. Malkan$^{5}$}
\author{Tommaso Treu$^{5,6}$}
\author{Vardha N. Bennert$^{7}$}
\author{Roberto J. Assef$^{8}$}
\author{Anna Pancoast$^{9,\ddag}$}

\affil{$^{1}$Korea Astronomy and Space Science Institute, Daejeon, 34055, Republic of Korea; \href{mailto:daeseongpark@kasi.re.kr}{daeseongpark@kasi.re.kr}}
\affil{$^{2}$National Astronomical Observatories, Chinese Academy of Sciences, Beijing 100012, China; }
\affil{$^{3}$Department of Physics and Astronomy, University of California, Irvine, CA 92697, USA; \href{mailto:barth@uci.edu}{barth@uci.edu}}
\affil{$^{4}$Astronomy Program, Department of Physics and Astronomy, Seoul National University, Seoul, 151-742, Republic of Korea; \href{mailto:woo@astro.snu.ac.kr}{woo@astro.snu.ac.kr}}
\affil{$^{5}$Department of Physics and Astronomy, University of California, Los Angeles, CA 90095, USA; \href{mailto:malkan@astro.ucla.edu}{malkan@astro.ucla.edu}, \href{mailto:tt@astro.ucla.edu}{tt@astro.ucla.edu}}
\affil{$^{6}$Department of Physics, University of California, Santa Barbara, CA 93106, USA}
\affil{$^{7}$Physics Department, California Polytechnic State University, San Luis Obispo, CA 93407, USA; \href{mailto:vbennert@calpoly.edu}{vbennert@calpoly.edu}}
\affil{$^{8}$N\'ucleo de Astronom\'ia de la Facultad de	Ingenier\'ia, Universidad Diego Portales, Av. Ej\'ercito Libertador	441, Santiago, Chile; \href{mailto:roberto.assef@mail.udp.cl}{roberto.assef@mail.udp.cl}}
\affil{$^{9}$Harvard-Smithsonian Center for Astrophysics, 60 Garden Street, Cambridge, MA 02138, USA; \href{mailto:anna.pancoast@cfa.harvard.edu}{anna.pancoast@cfa.harvard.edu}}

\altaffiltext{$\dag$}{EACOA fellow}
\altaffiltext{$\ddag$}{NASA Einstein fellow}
\altaffiltext{$\star$}{Based on observations made with the NASA/ESA Hubble Space Telescope, obtained at the Space Telescope Science Institute, which is operated by the Association of Universities for Research in Astronomy, Inc., under NASA contract NAS 5-26555. These observations are associated with program GO-12922.}

\begin{abstract}
We provide an updated calibration of \CIV\ $\lambda1549$ broad emission line-based single-epoch (SE) 
black hole (BH) mass estimators for active galactic nuclei (AGNs)
using new data for six reverberation-mapped AGNs at redshift $z=0.005-0.028$ 
with BH masses (bolometric luminosities) 
in the range $10^{6.5}$--$10^{7.5}$ \msun\ ($10^{41.7}$--$10^{43.8}$ \ergs).
New rest-frame UV-to-optical spectra covering 1150--5700 \AA\
for the six AGNs were obtained with the {\it Hubble Space Telescope (HST)}.
Multi-component spectral decompositions of the \emph{HST} spectra were
used to measure SE emission-line widths for the \CIV, \MgII, and
\Hb\ lines as well as continuum luminosities in the spectral region
around each line.
We combine the new data with similar measurements for a previous
archival sample of 25 AGNs to derive the most consistent and accurate
calibrations of the \CIV-based SE BH mass estimators against the
\Hb\ reverberation-based masses, using three different measures of
broad-line width: full-width at half maximum (FWHM), line dispersion
($\sigma_{\rm line}$) and mean absolute deviation (MAD). The newly
expanded sample at redshift $z=0.005-0.234$ covers a dynamic range in BH mass 
(bolometric luminosity) of $\log\ M_{\rm BH}/M_{\sun} = 6.5-9.1$ 
($\log\ L_{\rm bol}/$\ergs$=41.7-46.9$), 
and we derive the new \CIV-based mass
estimators using a Bayesian linear regression analysis over this
range.
We generally recommend the use of $\sigma_{\rm line}$ or MAD rather than FWHM
to obtain a less biased velocity measurement of the \CIV\ emission line,
because its narrow-line component contribution is difficult to
decompose from the broad-line profile.
\end{abstract}


\section{INTRODUCTION} \label{sec:intro}
\setcounter{footnote}{0}

Understanding the cosmic growth of the supermassive black hole (BH)
population and the coevolution of BH with their host galaxies 
is now recognized to be one of the essential ingredients for a complete picture of galaxy formation and evolution 
(see \citealt{FerrareseFord2005} and \citealt{KormendyHo2013}).
To probe the high-redshift BH population, and the evolution of the BH-galaxy scaling relations over cosmic time, 
it is essential to have reliable methods to determine black hole
masses in distant active galactic nuclei (AGN)  (\citealt{Shen2013}).

The rest-frame UV \CIV$\lambda1549$ broad emission line is commonly used for BH mass estimates in high-redshift AGNs 
(i.e., $2\lesssim z\lesssim 5$) when single-epoch (SE) optical spectra are available.
The method of deriving SE mass estimates based on broad-line widths
and continuum luminosities in quasar spectra relies on
reverberation-mapped (RM) AGN for its fundamental
calibration.\footnote{See a recent review by \citet{Bentz2016} on the
  current status and future prospects for RM studies.}  Achieving an
accurate calibration of \CIV-based SE BH mass estimators using the
most reliable AGN BH mass estimates obtained from RM is thus important
for improving the precision and accuracy of SE mass estimates for AGNs.
Due to a lack of direct \CIV\ RM measurements, however,
the \CIV\ SE calibration has been performed against the \Hb\ RM-based 
BH masses, which is the best practical approach at present. 
Note that the \Hb\ is so far the most studied and understood 
emission line in RM studies with many reliable \Hb-based RM results, 
which can thus be arguably regarded as the most reliable line for AGN 
BH mass measurements (see \citealt{Shen2013} for a related discussion).

Previously, \citet[][hereafter VP06]{VestergaardPeterson2006} have
provided a calibration of \CIV-based BH mass estimators using a sample
of low-redshift AGNs for which both \Hb\ RM measurements and
rest-frame UV spectra were available.  Since then, the number of AGNs
with BH mass estimates from RM has increased, as has the number of
AGNs for which \emph{Hubble Space Telescope (HST)} UV spectra have 
been obtained by the \emph{HST}. \citet[][hereafter
  P13]{Park+2013} have revisited the calibrations of \CIV-based BH
mass estimators by taking advantage of high-quality \HST\ UV spectra
for the reverberation-mapped AGN sample and using improved measurement
methods.  The P13 sample included 25 AGNs, of which six have estimated
$\mbh < 10^{7.5}$ \msun\ and only one has $\mbh < 10^{7.0}$ \msun. In
order to improve the calibration of SE BH masses at the low end
of the AGN mass range ($\mbh \lesssim 10^{7.5}$ \msun), it is
important to further expand the sample of AGNs having both RM
measurements and \HST\ UV spectroscopy.  Similarly, a calibration of
BH masses based on the broad \MgII\ $\lambda2798$  emission line,
another commonly used rest-frame UV line at intermediate redshifts,
has also been performed (see, e.g.,
\citealt{McLureJarvis2002,Wang+2009}). As with \CIV, there is also
much room for improvement in the calibration
of \MgII-based BH masses, and extending the calibration to a larger
sample of AGNs over a wider dynamic range in BH mass is a high priority.

There have been several efforts in the literature to improve 
the calibration of \CIV-based SE BH mass estimators, e.g., 
by taking advantages of the ratio of UV to optical continuum luminosities
(color dependence; \citealt{Assef+2011}), the ratio of FWHM to \linedisp\ 
of \CIV\ (line shape; \citealt{Denney2012}), the peak flux ratio of the 
$\lambda 1400$ feature to \CIV\ (Eigenvector 1; \citealt{Runnoe+2013, 
Brotherton+2015}), and the \CIV\ blueshift (\citealt{ShenLiu2012,Coatman+2017}).

As an extension of our previous work (P13), this paper presents new
\HST\ UV and optical spectra of six reverberation-mapped AGNs with BH
masses of $10^{6.5}$ to $10^{7.5}$ \msun. High quality spectra,
quasi-simultaneously covering the \CIV\ to \Hb\ spectral regions with
a consistent aperture size and slit width, were obtained with the
Space Telescope Imaging Spectrograph (STIS). The new data enable a
consistent comparison between the broad emission lines while
minimizing measurement systematics due to time variability or aperture
effects.

Using the new spectra, we provide updated calibrations of \CIV-based
SE BH mass estimators for three different measures of broad-line
width: the full-width at half maximum (FWHM) and the line dispersion
(\linedisp), which have been commonly used in previous work on SE mass
estimates, and the mean absolute deviation (MAD), which was recently
suggested by \citet{Denney+2016} to be a useful linewidth measure for
virial mass estimation.

We use a Bayesian linear regression method, which is independently
implemented for this work, to carry out the calibration of the
\CIV\ virial mass relation.
Our method follows the work  of \citet{Kelly+2012} (see also \citealt{Kelly2007})
using the \texttt{Stan} probabilistic programming language (\citealt{stan2015}).
The Bayesian methodology and model specifications for the
linear regression analysis will be described in detail in a
forthcoming paper (Park 2017, in preparation).

This paper is organized as follows.  In \autoref{sec:sample}, the
calibration sample, \HST\ observations, and data reduction procedures
are described.  In \autoref{sec:meas}, we present measurements of the
\CIV, \MgII, and \Hb\ emission lines and comparisons of their
profiles.  The new calibration of the SE virial mass estimators based
on the FWHM, \linedisp, and MAD of the \CIV\ line profile are
presented in \autoref{sec:calib} with a comparison to previous
calibrations and a test of methodological differences in the linear
regression analysis. We also systematically compare our 
updated calibration with the corrected prescriptions in the literature 
for the \CIV\ BH mass calibration described above.
We summarize this work and provide discussions
in \autoref{sec:summary}.
The following standard cosmological parameters were adopted to calculate distances: 
$H_0 = 70$~km~s$^{-1}$~Mpc$^{-1}$, $\Omega_{\rm m} = 0.30$, and $\Omega_{\Lambda} = 0.70$, 
which is the same as used by P13.

\section{Sample, \HST\ Observations, And Data Reduction} \label{sec:sample}

The sample for this work is based on the sample of 25 AGNs 
(BH mass $\log M_{\rm BH}/M_{\odot} = 7.0-9.1$, 
bolometric luminosity\footnote{The bolometric luminosity is computed as 
$L_{\rm bol}=3.81 \times \lambda L_{1350}$ 
(see \citet{Shen+2008} and references therein).} $\log\ L_{\rm bol}/$\ergs$=43.2-46.9$, 
redshift $z=0.009-0.234$) from P13,
supplemented by six new AGNs at redshift $z=0.005-0.028$ that have low-mass BHs 
(i.e., $\log M_{\rm BH}/M_{\odot} = 6.5-7.5$) from \Hb-based RM measurements
and low bolometric luminosities (i.e., $\log\ L_{\rm bol}/$\ergs$=41.7-43.8$). The
P13 sample contains reverberation-mapped AGNs with available archival
\HST\ spectra, selected by taking into account data quality, spectral
coverage, and contamination of \CIV\ by absorption features. The
enlarged dynamic range in mass for the expanded sample enables us to
calibrate the \CIV\ SE virial relationship over almost three orders of
magnitude in BH mass.  The new targets have been selected from recent
RM programs.  These include Arp 151, Mrk 1310, NGC 6814, and SBS
1116+583A from the Lick AGN Monitoring Project 2008 campaign
(\citealt{Bentz+2009,Park+2012}), Mrk 50 from the Lick AGN Monitoring
Project 2011 campaign (\citealt{Barth+2011:mrk50}), and the
Kepler-field AGN Zw 229-015 (\citealt{Barth+2011:zw229}).
\autoref{tab:RMdata} summarizes the properties of the P13 AGN sample
and the six new objects presented in this work.  Note that the virial
factor $f$ with its uncertainty is adopted from \citet{Park+2012ApJS}
and \citet[][see also \citealt{Woo+2013,Woo+2015}]{Woo+2010} and applied to all RM BH masses (i.e., $\log f =
0.71 \pm 0.31$), which is consistent with previous measurements and
also with direct measurements by
\citet{Pancoast+2012,Pancoast+2014-II}.
The $f$ represents the dimensionless scale factor of order unity 
that depends on the detailed geometry, kinematics, and inclination of broad-line region (BLR),
which is thus used to convert measured virial product into actual black hole mass ($M_{\rm BH}=f \times \rm VP_{\rm BH}$).
The adopted uncertainty (0.31 dex) for the virial factor is derived from the scatter of the AGN \msigma\ relation (0.43 dex),
which gives an upper limit of random scatter of the virial factor itself after subtracting off in quadrature 
the assumed intrinsic scatter (0.3 dex) of the relation (see also the related discussion in \citealt{Park+2012}).
Note that the virial factor uncertainty is the dominant portion of the error 
budget for the RM masses, since the measurement uncertainty propagated 
from the reverberation lags and \Hb\ line widths is substantially smaller 
than this 0.31 dex uncertainty for individual AGNs (see \autoref{tab:RMdata}).

For the six new AGNs, we obtained UV spectra with STIS as part of
\HST\ program GO-12922 (PI: Woo).  In addition to the UV data, optical
spectra were also obtained quasi-simultaneously (during the same
\HST\ visit) with a consistent slit width and aperture size. 
Note that temporal gaps between the end of optical exposures and the start of UV exposures were less than $\sim 6$ minutes.
Individual exposures in and between UV gratings were obtained within a maximum temporal gap of $\sim50$ minutes.
The ability to obtain nearly simultaneous UV and optical spectra through a
consistent aperture is a unique capability of the STIS instrument, and
is essential in order to minimize possible systematic biases from AGN
variability and different amounts of host galaxy and narrow-line
region contributions.

We used the G140L, G230L, and G430L gratings with the \texttt{52x0.2}
slit (i.e., a long slit of width 0\farcs2) to acquire a spectrum covering
the Ly~$\alpha$, \CIV, \MgII, and \Hb\ emission lines for each
target. The consistent and small spectroscopic aperture has the
benefit of minimizing the contamination from host-galaxy
starlight. For the CCD G430L observations, we used the E1 aperture
position to minimize losses due to the imperfect charge transfer
efficiency as recommended in the STIS instrument handbook.  Total
integrations of 1170--1464 s for G140L, 627--1471 s for G230L, and
120--200 s for G430L respectively were split into two or three
exposures depending on the grating, and dithered along the slit for
optimal cleaning of cosmic-ray hits and bad pixels.
The observations are summarized in \autoref{tab:ObsDetailes}.
Note that the slit PA was not constrained, in
order to maximize the \HST\ scheduling opportunities.
But the three grating data for each object were obtained in a single \HST\ 
visit with the same orientation.

While we used the fully reduced data provided by the \HST\ STIS pipeline for the UV gratings,
we performed a custom reduction for the optical grating data from the raw science and reference files 
in order to improve the cleaning of cosmic-ray charge transfer trails in raw
images from the
badly degraded STIS CCD. 
Based on the standard reduction of the STIS pipeline, an additional
cosmic-ray removal step was added to the processes employing the
\texttt{LA\_COSMIC} \citep{vanDokkum2001} routine following the
approach described by \citet{Walsh+2013}.
The raw data for the optical G430L grating were first calibrated with
the \texttt{BASIC2D} task including trimming the overscan region, bias and dark subtraction, and flat-fielding. 
Cosmic-rays and hot pixels were then cleaned with \texttt{LA\_COSMIC}, and wavelength calibration was performed.

\begin{figure*} 
	\centering
	\includegraphics[width=0.95\textwidth]{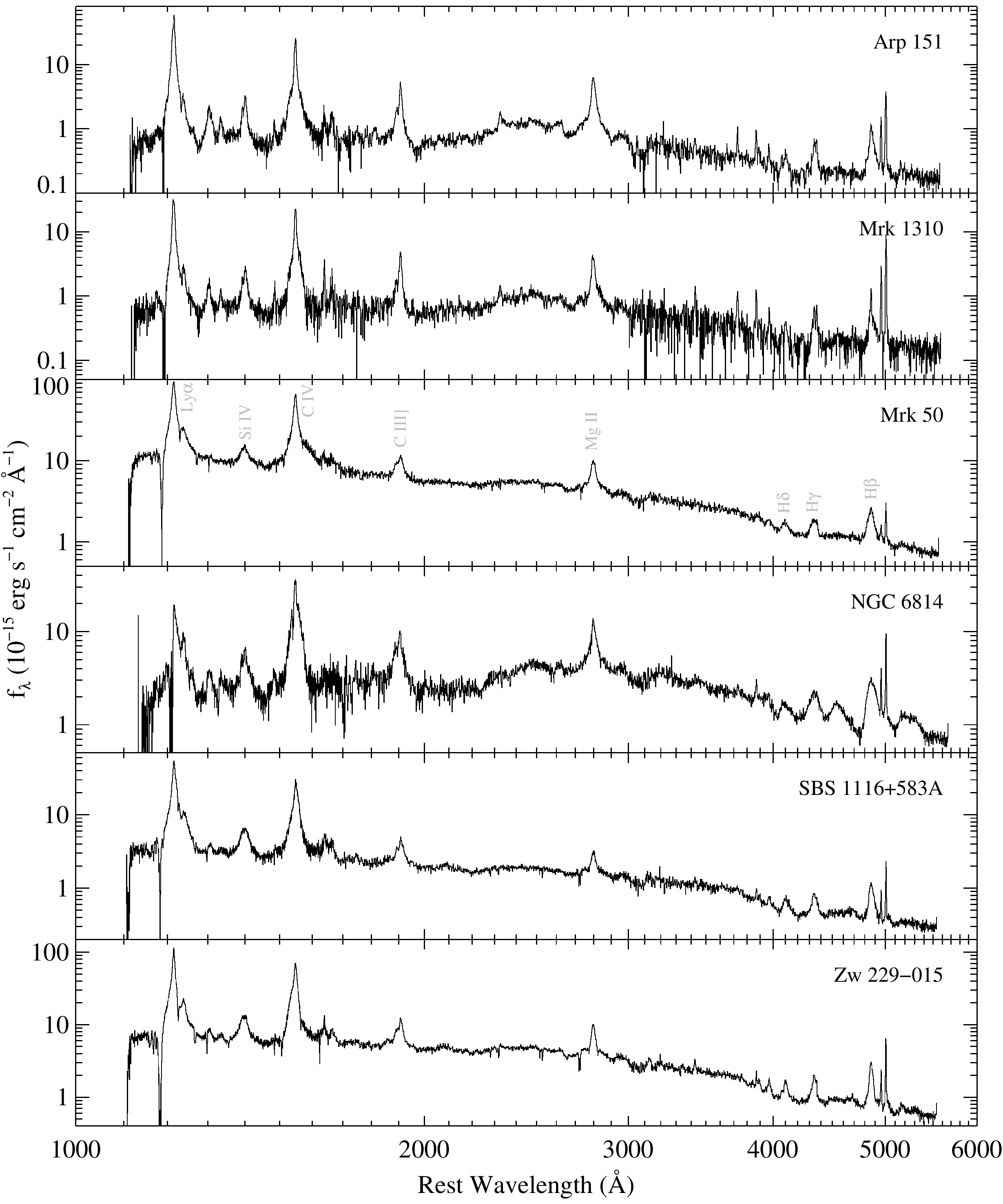} 
	\caption{
		Final fully reduced and combined STIS spectra for our sample of the six low-mass AGNs.
	}
	\label{fig:specall}
\end{figure*}

The dithered individual exposures for each grating were then aligned
and combined using the \texttt{IMSHIFT} and \texttt{IMCOMBINE} PYRAF
tasks.  After that, one-dimensional spectra from each grating were
extracted with the \texttt{X1D} task and then joined together to
produce a final single spectrum by taking into account the flux and
noise levels in overlapping regions around $\sim1700$ \AA\ and
$\sim3100$ \AA.
Following P13, we corrected the spectra for Galactic extinction using
the values of $E(B-V)$ from \citet{Schlafly+2011} as listed in the
NASA/IPAC Extragalactic Database (NED), and the reddening curve of
\citet{Fitzpatrick1999}.  \autoref{fig:specall} shows the fully
reduced and calibrated rest-frame spectra of the six AGNs.

\section{Spectral Measurements} \label{sec:meas}

To measure the broad emission-line widths and the continuum luminosity
adjacent to each broad line, we carried out a multi-component spectral
decomposition analysis to the spectral region surrounding
\CIV\ $\lambda1549$, \MgII\ $\lambda2798$, and \Hb\ $\lambda4861$.  A
combination of these two observables, line width and continuum
luminosity measured from a single-epoch spectrum, is commonly used to
estimate BH masses via the SE BH mass estimators because they can be
adopted as reasonable proxies for velocity of the broad-line gas
clouds and the size of BLR 
\citep{Kaspi+2000,Kaspi+2005,Bentz+2006,Bentz+2009RL,Bentz+2013}, respectively.
Following the standard
approach that has been adopted in previous works 
\citep[e.g.,][]{Shen+2008,Shen+2011}, we measure monochromatic
continuum luminosities at $1350$ \AA, $3000$ \AA, and $5100$ \AA\ to
compute SE virial masses from \CIV, \MgII, and H$\beta$, respectively.

\begin{figure*} 
	\centering
	\includegraphics[width=0.95\textwidth]{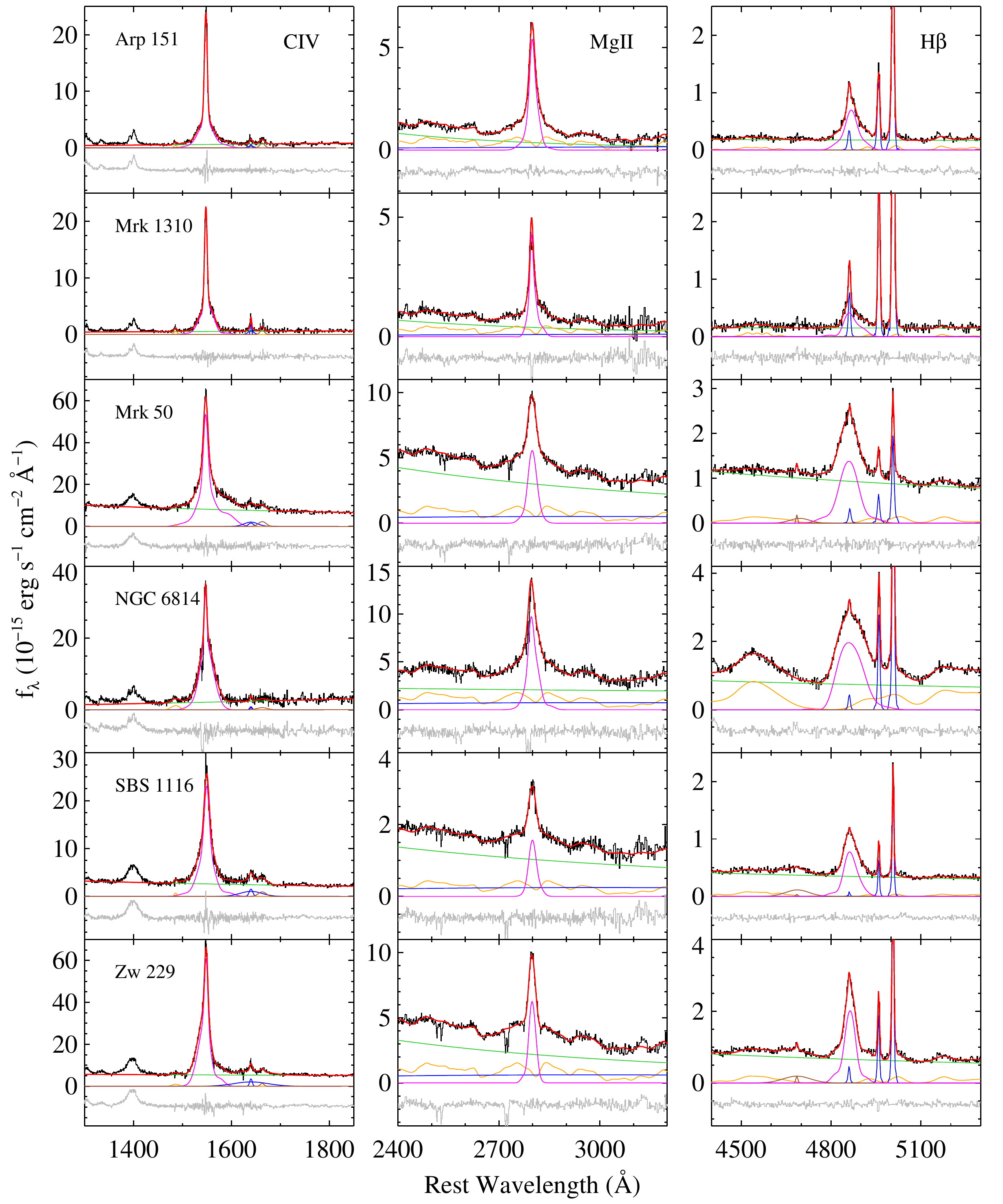} 
	\caption{
		Multi-component spectral decompositions in  the spectral regions of three major broad emission lines, 
		\CIV\ $\lambda1549$, \MgII\ $\lambda2798$, and \Hb\ $\lambda4861$, for our 6 AGNs.
		In each panel, the observed spectrum (black) is decomposed into various components. 
		\textit{Left (\CIV)}: the power-law continuum (green), \CIV\ $\lambda1549$ (magenta), and other nearby blended lines, 
		\ion{N}{4}] $\lambda1486$ (orange), \ion{He}{2} $\lambda1640$ (blue), \ion{O}{3}] $\lambda1663$ (brown). 
		\textit{Middle (\MgII)}: the power-law continuum (green), \FeII\ template (orange), Balmer continuum (blue), 
		\MgII\ $\lambda2798$ (magenta).
		\textit{Right (\Hb)}: the power-law continuum (green), \FeII\ template (orange), three narrow emission lines (\Hb, [\ion{O}{3}] $\lambda\lambda4959, 5007$; blue), broad \Hb\ (magenta), and the broad and narrow \ion{He}{2} $\lambda4686$ components (brown; only included if blended with \Hb).
		The red line in each panel indicates the full model combining all the best-fit model components. 
		Each bottom gray line represents residual, i.e.,  data $-$ model, shifted downward for clarity.
	}
	\label{fig:modelfitall}
\end{figure*}

Our fits are based on a local decomposition of the spectral region
around each broad line, rather than a global decomposition of the
entire UV-optical spectrum. Owing to the complexity of the spectra and
the large number of emission-line and continuum components that are
present, we found that local decompositions are able to achieve a more
precise fit to the data around each line than would be possible in a
simultaneous, global fit to the full STIS spectrum
(see later \autoref{sec:meas:compare} for a discussion on the global versus local fits).
The local
spectral decomposition technique employed here is based on those by
P13 and \citet{Park+2015} and slightly updated and modified for the
STIS data and the spectral region in question.
Our spectral modeling method consists of separate procedures for
continuum fitting and line emission fitting, applied independently to
the \CIV, \MgII, and H$\beta$ regions of the data.
During fitting, model parameters are optimized using \texttt{mpfit}
\citep{Markwardt2009} in IDL.  The model components and fitting
details for each of the \Hb, \MgII, \CIV\ line regions are described
in the following subsections, and the decomposition results are given
in \autoref{fig:modelfitall}.

\subsection{\Hb}

We used the multi-component spectral decomposition code developed by 
\citet{Park+2015} for modeling the \Hb\ region of our STIS data.
In brief, the code works by first simultaneously fitting a pseudocontinuum 
that consists of a single power-law, an \FeII\ template, and a host-galaxy 
template in the surrounding continuum regions of $4430-4770$ \AA\ and 
$5080-5450$ \AA, and then fitting the \Hb\ emission line complex with 
Gauss-Hermite series functions \citep{vanderMarelFranx1993,Cappellari+2002} 
for one broad emission component (\Hb) and three narrow emission components 
(\Hb, [\ion{O}{3}] $\lambda\lambda4959, 5007$), and two Gaussian functions 
for the nearby blended \ion{He}{2} $\lambda4686$ emission line 
after subtracting the best-fit pseudocontinuum model
(see \citealt{Park+2015} and references therein for details of the measurement procedure) (see also \citealt{Woo+2006,Bennert+2015,Runco+2016}). 
The \Hb\ line widths, ${\rm FWHM}_{\rm H\beta}$ and $\sigma_{\rm
  H\beta}$, are measured from the best-fit broad line model (i.e., the
Gauss-Hermite series function), and the continuum luminosity at $5100$ \AA, $\lambda L_{5100 \text{\rm \AA}}$, is measured from the best-fit power-law model.

Note that there are two differences between the method adopted here and the approach given by \citet{Park+2015}, 
specifically in the model components used for the \FeII\ emission and host-galaxy starlight.
The template for host-galaxy starlight is excluded in this work
because stellar absorption features, which is critical to 
achieve reliable host galaxy template fits, are not observable in the
small-aperture STIS spectra. The minimal contribution of host-galaxy
light and the relatively low signal-to-noise ratio and spectral
resolution of the STIS optical data make it difficult to detect any
host galaxy features in the optical data.  Moreover, the fits did
not converge when we included the host-galaxy starlight component in
the model.
As a rough check, 
we provide a crude estimate of an upper limit for host starlight 
contribution to the STIS spectra using the object, SBS 1116+583A, 
which shows the highest host galaxy fraction in the ground-based 
spectroscopic observations (see \citealt{Park+2012,Barth+2015}) from our STIS sample.
The host galaxy flux in the STIS spectrum can be roughly estimated by 
subtracting the AGN flux at $5100$ \AA, which is obtained by subtracting 
off the \HST\ imaging-based galaxy flux at $5100$ \AA\ \citep{Bentz+2013} from the 
ground-based spectroscopic total flux at $5100$ \AA\ \citep{Park+2012}, 
from the total flux at $5100$ \AA\ of the STIS spectrum.
The resulting host galaxy fraction in the STIS spectrum is found to be $\sim31$\%.
Note that the other AGNs will have much lower contributions than this
due to the lower host galaxy fractions shown by \citet{Park+2012} and \citet{Barth+2015}.

Available \FeII\ templates for the \Hb\ region include empirically
constructed monolithic templates by \citet{BorosonGreen1992} and
\citet{Veron-Cetty2004}, a theoretical template by
\citet{BruhweilerVerner2008}, and a semi-empirical multi-component
template by \citet{Kovacevic+2010}.  After performing extensive tests
using each of the templates and a linear combination of the templates
for our STIS data, we opted to use the template by
\citet{Kovacevic+2010} based on its overall performance as quantified
by the $\chi^2$-statistics and residuals of the fits (see also
\citealt{Barth+2013,Barth+2015}).  As expected, the multi-component
template generally performs better than monolithic templates,
particularly for the objects showing strong \FeII\ emission.  The
\citet{Kovacevic+2010} template appears to be the best currently
available for accurately fitting diverse \FeII\ emission blends in
AGNs, by allowing for different relative intensities between five
\FeII\ multiplet subgroups.
To sum up, we follow the method described by \citet{Park+2015} except
that we used the template of \citet{Kovacevic+2010} instead of
\citet{BorosonGreen1992} for \FeII\ emission, and we omitted the
host-galaxy starlight template from the fits.

\subsection{\MgII}

For the \MgII\ spectral region, we first fit a pseudocontinuum model 
in the surrounding continuum regions of $2450-2750$ \AA\ and $2850-3100$ \AA. 
The pseudocontinuum model is composed of 
a single power-law function representing the AGN featureless continuum, 
an \iron\ emission template, 
and an empirical model for the Balmer continuum.
We adopt the UV \iron\ template from \citet{Tsuzuki+2006}, which is made from observations of I Zw 1.
Using the template of \citet{Tsuzuki+2006} is arguably better for
modeling the \MgII\ line region than using that of
\citet{VestergaardWilkes2001}, because it contains semi-empirically
constrained \FeII\ contribution underneath the \MgII\ line, while the
template by \citet{VestergaardWilkes2001} has no \FeII\ flux at all
under the \MgII\ line due to the difficulty of decomposing this
spectral region.

Based on the investigations of \citet{Grandi1982} and \citet{Wills+1985}
(see also \citealt{MalkanSargent1982} for the first practical measurement of the Balmer continuum shape),
\citet{Dietrich+2002,Dietrich+2003} described a practical procedure for the Balmer continuum modeling in high-$z$ quasar spectra, 
which has become a standard practice for fitting the Balmer continuum.
This empirical model assumes that the Balmer continuum is generated from partially optically thick gas clouds with uniform effective temperature ($T_{e}=15,000$ K) as 
\begin{equation}
F_{\lambda}^{\rm BaC}(A,T_{e},\tau_{\rm BE})=A B_{\lambda} (T_{e})\left( 1 - e^{-\tau_{\rm BE}(\lambda/\lambda_{\rm BE})^3} \right), \lambda \le \lambda_{\rm BE},
\end{equation}
where $A$ and $\tau_{\rm BE}$ are normalized flux density and optical
depth at the Balmer edge ($\lambda_{\rm BE}=3646$ \AA), and $B_{\lambda} (T_{e})$ is the Planck function at the electron temperature $T_{e}$. At $\lambda > \lambda_{\rm BE}$, higher order Balmer lines using the relative intensity calculations from \citet{StoreyHummer1995} are used to represent the smooth rise to the Balmer edge. 
Many studies (e.g., \citealt{Kurk+2007, Wang+2009, Greene+2010,
  DeRosa+2011,DeRosa+2014, Ho+2012, ShenLiu2012, Kokubo+2014}) have
used variants of this method with slightly different ways of
constraining the model parameter ranges based on the available data
quality and spectral coverage.
We found that if we treat all three parameters in the model ($A,T_{e},\tau_{\rm BE}$) as free parameters during fitting as done by \citet{Wang+2009} and \citet{ShenLiu2012},
they were very poorly constrained due to the degeneracy with the power-law continuum and \iron\ emission blends.

Recently, \citet{Kovacevic+2014} suggested an improved way to
constrain the normalization $A$, by taking into account a fact that
$A$ can be obtained by calculating the sum of all intensities of
higher order Balmer lines at the Balmer edge.  We followed this
procedure with slight modifications.  The value of $A$ was separately
determined from the intensity calculation using high order Balmer
lines (up to 400) with the template line profile adopted from the
best-fit \Hb\ emission line model obtained above.  The other
parameters, $T_{e}$ and $\tau_{\rm BE}$, were then fitted
simultaneously with the \FeII\ template and power-law function during
the pseudocontinuum modeling.  Note, however, that following the
Dietrich approach, the temperature was finally fixed to be $15,000$ K
and optical depth was allowed to vary between 0.1 and 2. We also
independently checked that the constrained Balmer continuum component
only exhibits marginal changes over temperature ranging from 10,000 K
to 30,000K and optical depth varying from 0.1 to 2.
Note that the resulting continuum luminosity estimates are 
consistent with each other within $\sim0.04$ dex scatter.

After subtracting the best-fit pseudocontinuum model,
the \MgII\ emission line was fitted using a linear combination of a
sixth-order Gauss-Hermite series and a single Gaussian function to
account for its full line profile, 
typically showing a more peaky core (i.e., narrower and sharper line peak) and
more extended wings than a Gaussian profile,
in the spectral region $\sim 2700 - 2900$ \AA.
We use the full line profile without a decomposition of narrow and broad components 
for line width measurements for UV lines in this work (the same approach adopted by P13), in contrast to \Hb, 
because no reliable and clear distinction between broad and narrow components in the UV lines is usually possible,
and sometimes no narrow components of the UV lines are seen at all,
although their presence is still uncertain and under debate.
Thus, the \MgII\ line widths, ${\rm FWHM}_{\rm Mg II}$ and $\sigma_{\rm Mg II}$, are measured from the best-fit full line profile, and continuum luminosity at $3000$ \AA, $\lambda L_{3000 \text{\rm \AA}}$, is measured from the best-fit power-law function.
During fitting, Galactic absorption lines such as \ion{Fe}{2}
$\lambda\lambda 2586,2600$, \ion{Mg}{2} $\lambda\lambda 2796,2803$,
and \ion{Mg}{1} $\lambda 2852$ (cf. \citealt{Savaglio+2004}) were
masked out with exclusion windows.

\subsection{\CIV}

Spectral measurements for the \CIV\ line region in the previous
archival sample of local 25 RM AGNs were described by P13.  Here we
focus on analysis of the 6 objects with newly obtained STIS data.  We
used the same methods as in P13 for consistency and to avoid
additional systematic biases.  We fit the AGN featureless continuum
with a single power-law function, and we chose to omit a UV iron
template (e.g., \citealt{VestergaardWilkes2001}) from the fits because
no clear contribution of iron emission over the \CIV\ region is observed.
Although we performed a test including the UV iron template in the model as in P13, 
its contribution was too small to be constrained accurately with the template at least in our sample (see also \citealt{Shen+2008,Shen+2011}).

After the best-fit continuum model is subtracted, the \CIV\ emission line is fitted with 
a linear combination of a sixth-order Gauss-Hermite series and a single Gaussian function.
The contaminating nearby blended emission lines (e.g., \ion{N}{4}] $\lambda1486$, \ion{Si}{2} $\lambda1531$, \ion{He}{2} $\lambda1640$, \ion{O}{3}] $\lambda1663$) are fitted simultaneously as well using up to two Gaussian functions for each line.
Again, we use the combined model of one Gauss-Hermite function and one Gaussian function to fit the full line profile of \CIV\ 
without decomposing it into broad and narrow components.
The \CIV\ line widths, ${\rm FWHM}_{\rm C IV}$ and $\sigma_{\rm C IV}$, are measured from the best-fit full line profile, and continuum luminosity at $1350$ \AA, $\lambda L_{1350 \text{\rm \AA}}$, is measured from the best-fit power-law function.
Narrow absorption spikes are masked out using a $3\sigma$ clipping
threshold during fitting, and broad absorption features around the
line center are masked out manually with exclusion windows
(see P13 and references therein for more details of the \CIV\ measurement method and results for the archival sample).

\subsection{Measurement uncertainty estimation} \label{sec:meas:err}

Uncertainties for the above spectral measurements are estimated with the Monte Carlo method used by 
\citet{Park+2012} and P13 (see also \citealt{Shen+2011,ShenLiu2012}).
For each spectral region, 1,000 mock spectra are generated by
resampling the original spectra with the addition of Gaussian random noise based on the error spectrum for each object.
We then measure line widths and luminosities from each of the mock spectra using the same measurement methods and
take the standard deviation of the distribution of the measurements as
the estimate of measurement uncertainty.

Typical uncertainty levels of line widths for all objects are found to
be $\sim 2-4$\% with a maximum of $\sim 17$\% due to the high quality
of the \HST\ spectra.  For continuum luminosity, we derive
uncertainties of $\sim 1-2$\% with a maximum of $\sim 6$\%.  These are
small compared to the overall systematic mass uncertainty of $\sim0.4$
dex in the SE virial method.  Covariances between the measurement
uncertainties of the line widths and luminosity for each object in a
logarithmic scale are also estimated from the resulting distributions
of the Monte Carlo simulations, which are given as ${\rm cov}(\log
\lambda L_{\lambda},\log {\rm FWHM})=\rho \sigma(\log \lambda
L_{\lambda}) \sigma(\log {\rm FWHM})$ and ${\rm cov}(\log \lambda
L_{\lambda},\log \sigma_{\rm line})=\rho \sigma(\log \lambda
L_{\lambda}) \sigma(\log \sigma_{\rm line})$
where ${\rm cov}$, $\rho$, and $\sigma$ are the covariance, correlation coefficient, and measurement uncertainty of the logarithms of the luminosity and line widths, respectively.
\autoref{tab:UVmeasurement} lists the line widths and luminosities for our sample along with the measurement uncertainties and their error correlation coefficients.

\subsection{Continuum luminosities and emission-line widths} \label{sec:meas:compare}

There are several issues in regard to measuring continuum luminosities and line widths accurately.
It is important to take into account the Balmer continuum over the
\MgII\ line region to accurately decompose the power-law continuum for
luminosity measurements.  Based on our investigation of the STIS data,
the $\lambda L_{3000 \text{\rm \AA}}$ values would be overestimated by
$\sim 0.14$ dex on average if the Balmer continuum component is not
accounted for, which is also consistent with the investigation of
\citet{ShenLiu2012}, who found $\sim 0.12$ dex systematic offset.
This bias will then be propagated into final \mbh\ estimates by as
much as a $\sim 0.07$ dex ($\sim 17$\%) systematic offset if the Balmer
continuum model is not included properly.

If the original \citet{VestergaardWilkes2001} \FeII\ template is used,
FWHM (\linedisp) estimates are overestimated by $\sim 0.03$ ($0.07$) dex on average 
compared with results derived using the template of \citet{Tsuzuki+2006}.
This will again be propagated into \mbh\ estimates by up to $\sim
0.06$ ($0.14$) dex of systematic offset,
which is consistent with the result of \citet{Nobuta+2012}.

To make the maximum use of the wide spectral coverage of our STIS
data, we have also performed extensive tests of global continuum fits
covering \CIV\ to \Hb\ simultaneously.  Using a more flexible double
power-law model to represent the AGN featureless continuum, we fit a
pseudocontinuum model including the Balmer continuum model and
\FeII\ emission to many line-free continuum regions (see also, e.g.,
\citealt{ShenLiu2012} and \citealt{Mejia-Restrepo+2016} for related
recent work).  The global continuum fits produced results consistent
with the local continuum fits described in the previous sections,
except that the global fits failed to constrain the \FeII\ emission on
the red side of the \MgII\ regions for Arp 151 and Mrk 1310.  This is
probably because the double power-law model is not flexible enough to
properly describe steep local slope changes around \MgII\
for these two objects, which is coming from intrinsic changes of
spectral shapes and/or from strong internal reddening, along with the
incompleteness of currently available UV \FeII\ templates across the
regions.  In any case, there is no significant improvement of the
global fits compared to the local fits.

\begin{figure*}[!ht]
	\centering
	\includegraphics[width=0.95\textwidth]{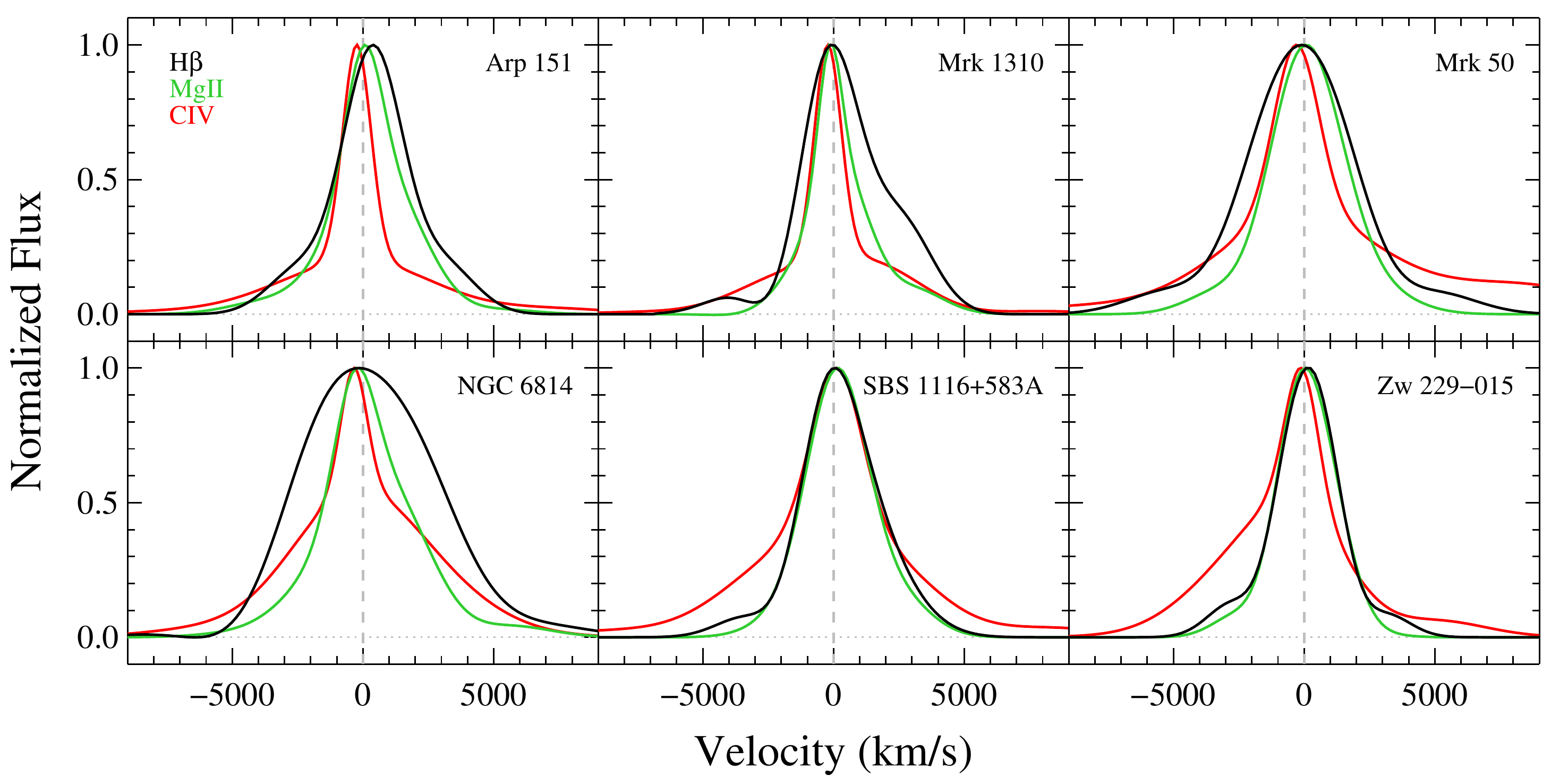} 
	\caption{
		Comparison of the modeled emission line profiles, normalized by each peak, of the \CIV\ (red), \MgII\ (green), and \Hb\ (black) for our  STIS sample.
	}
	\label{fig:comp_profiles}
\end{figure*}

\begin{figure*}[!ht]
	\centering
	\includegraphics[width=0.95\textwidth]{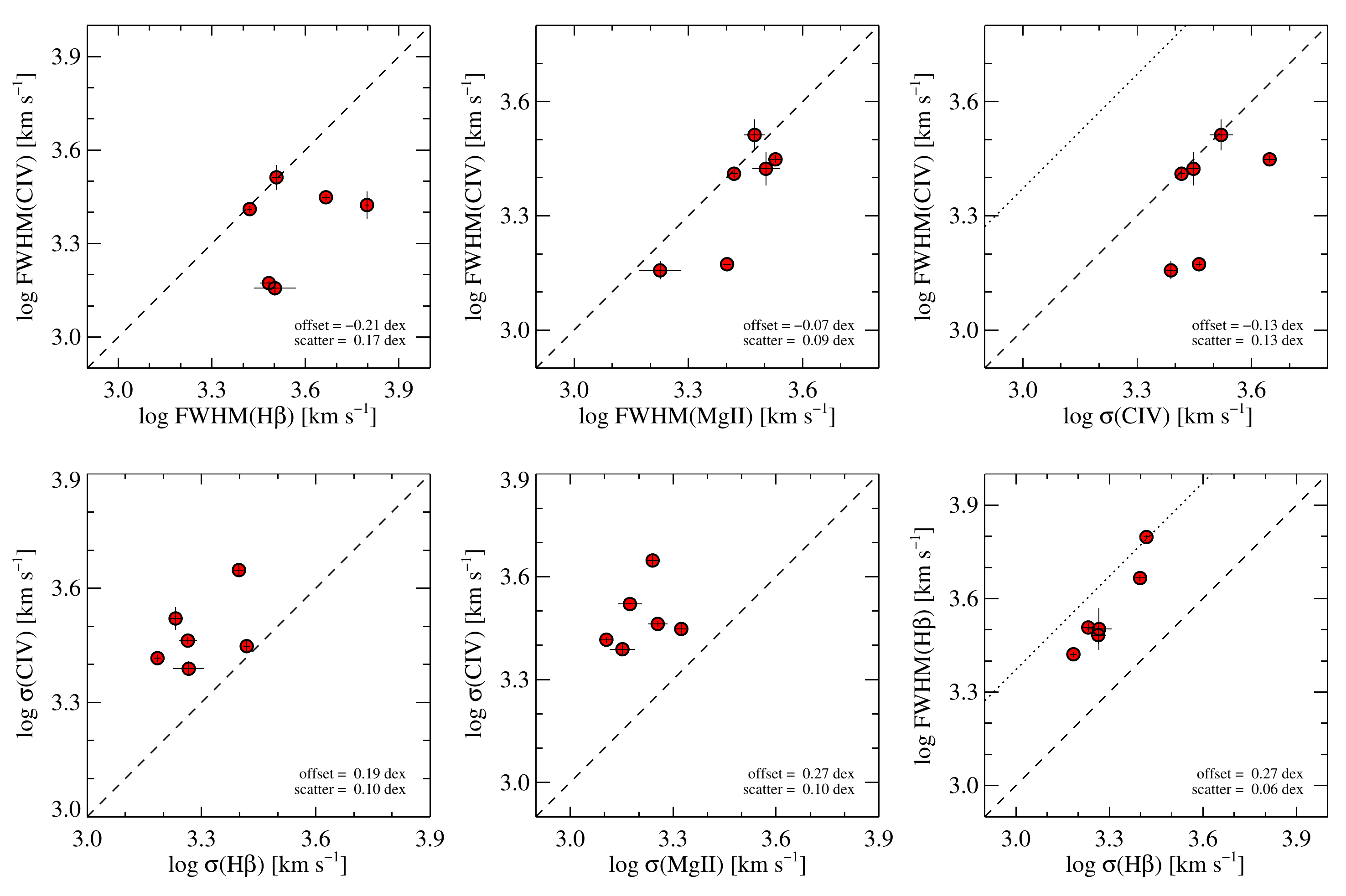} 
	\caption{
		Intercomparison of line width measurements, FWHM and \linedisp, of the \CIV, \MgII, and \Hb\ for our STIS sample.
		The dashed line represents a one-to-one relation in each panel. 
		The dotted line shows the ratio of FWHM to \linedisp\ for a Gaussian profile.
		The mean offset (i.e., average of line width differences) and 1$\sigma$ scatter (i.e., standard deviation of line width differences) are given at the lower right corner in each panel.
	}
	\label{fig:comp_linewidths}
\end{figure*}

The simultaneous coverage of our STIS data also makes it possible to consistently compare the major UV and optical emission lines (\CIV, \MgII, \Hb)
without biases from intrinsic variability (\autoref{fig:comp_profiles}). 
The \CIV\ profile shows on average more peaky cores with extended
wings than those of the \MgII\ and \Hb\ lines (see also \citealt{Wills+1993,Brotherton+1994}).
There is no significant velocity offset between the line peaks of all the three emission lines.
\autoref{fig:comp_linewidths} compares FWHM and \linedisp\ line width measurements among the emission lines. 
Although it is hard to draw a clear picture due to small-number statistics,
we  use only our STIS sample of six AGNs 
in order to perform a consistent comparison between the line widths in
quasi-simultaneously observed data.  We find that the FWHM of \CIV\ is on
average smaller than that of \Hb\ (and \MgII) FWHM, which may indicate that
FWHM for \CIV\ is not a good proxy for virial BLR velocity, probably
due to contamination from a non-reverberating \CIV\ core component
\citep{Denney2012}.
This contamination would be one of the biases correlated with 
Eigenvector 1 (EV1; \citealt{BorosonGreen1992}) as discussed by 
\citet{Runnoe+2013,Runnoe+2014} and \citet{Brotherton+2015} 
who investigated and used the peak flux ratio of the $\lambda1400$ feature to 
\CIV, as a UV indicator of the EV1, to correct for the \CIV-based BH masses.
However, the interpretation is not straightforward. It could also be the case
that the BLR geometry is different for the regions emitting these
three lines, resulting in different individual
virial factors ($f$) for each line (see also \citealt{Runnoe+2013orientation}).
On the other hand, $\sigma_{\rm C IV}$ is on average larger than $\sigma_{\rm H\beta}$,
which is consistent with the simple virial expectation of stratified
BLR structure and the shorter reverberation lags of \CIV\ (\citealt{PetersonWandel1999,Kollatschny2003}), 
thus corroborating the use of \linedisp\ over  FWHM for \CIV-based BH mass estimates.
More detailed intercomparisons and systematic investigation of
multi-line properties including more objects from the literature 
will be presented in a forthcoming paper.

\section{Bayesian Calibration of \CIV-based \mbh\ estimators} \label{sec:calib}

Now that we have the continuum luminosity and line width measurements from single-epoch spectra,
we can perform a calibration of the \CIV-based SE BH mass estimators against the \Hb\ 
RM-based BH masses as fiducial baseline. We assume that the BH masses
from \Hb\ RM are the most reliable mass estimates available for these
galaxies, and our goal is to find the combination of the SE measurements
that most closely reproduces the  RM mass scale, using the following equation:
\begin{eqnarray}\label{eq:calib}
\log \left(\frac{M_{\rm BH}^{\rm RM} }{M_\odot}\right)
&~=~&  \alpha ~+~ \beta ~ \log\left(\frac{\lambda L_{1350{\rm \text{\AA}}}^{\rm SE} }{10^{44}~\rm erg~s^{-1}}\right) \nonumber\\
&~+~&  \gamma ~ \log \left(\frac{\varDelta V_{\rm CIV}^{\rm SE}}{1000~ \rm km~s^{-1}}\right),
\end{eqnarray}
where $\varDelta V_{\rm CIV}^{\rm SE} = \rm FWHM^{\rm SE}_{\rm CIV}$
or $\sigma^{\rm SE}_{\rm CIV}$.
This equation essentially expresses the virial relation
$M_\mathrm{BH} \sim r_\mathrm{BLR} V^2/ G$, assuming that BLR radius
scales with AGN luminosity according to $r_\mathrm{BLR}\propto
L^\beta$ and allowing for the virial exponent $\gamma$ to differ from the
physically expected value of 2 in order to achieve the best fit.

Note that calibrations of the BH mass estimators based on the emission lines which have no direct reverberation measurements, 
e.g., \CIV\ and \MgII, have been performed indirectly against the \Hb-based reverberation results for the same objects if available (e.g., VP06 and P13).
Although there are a few direct (or in some cases tentative)
reverberation results for the  UV emission lines 
\citep[e.g.,][see also \citealt{Shen+2016}]{Kaspi+2007,Trevese+2014},
most of the available reverberation studies have been done with the
\Hb\ line, which gives the most reliable AGN BH masses at present.

To perform the calibration, we adopt a Bayesian approach to linear regression analysis. 
An advantage of the Bayesian method over the traditional $\chi^2$-based is that 
by obtaining probability density functions (PDFs) for parameters of interest instead of just calculating a point estimate,
it provides more reliable uncertainty estimates, incorporating all the error sources modeled and simply marginalizing over nuisance parameters. It is also easy to explore covariance between parameters from resulting joint probability distributions.
Bayesian linear regression method outperforms other previous classical methods especially when the measurement error of independent variables is large and/or the sample size is small (see \citealt{Kelly2007}).

For full Bayesian inference, we use the \texttt{Stan} probabilistic programming language \citep{stan2015}, 
which contains an adaptive Hamiltonian Monte Carlo (HMC; \citealt{Neal2012,HMC2013}) No-U-Turn
sampler (NUTS; \citealt{NUTS2014}) as its sampling engine. This provides a simple implementation for specifying complex hierarchical Bayesian models
and achieves good computational efficiency.
We set up the Bayesian hierarchical model following \citet{Kelly+2012} and implement it by referring to \citet{BDA3rd, dBDA2nd, stan-manual2015}.
The practical details of the sampler 
and the model specification will be described in a separate paper
describing the methodology (Park 2017, in preparation).
As a brief summary, 
the $t$ distribution is adopted to obtain outlier-robust statistical inference 
following the investigation of \citet{Kelly+2012} with an additional 
improvement of treating degrees-of-freedom in the $t$ model as a free parameter, 
instead of fixing it to be a pre-selected constant.
Thus, the likelihood function, which is specifying the measurement, 
regression, and covariate distribution models, is built with the $t$ distributions, 
and the prior distributions are specified based on the suggestions by 
\citet{Barnard+2000,BDA3rd,dBDA2nd}.
Our Markov chain Monte Carlo (MCMC) simulations have been run 
via the \texttt{PyStan} package (v2.9.0; \citealt{pystan290}) with 
careful assessment of the convergence of the MCMC chains.

\subsection{The single-epoch mass calibration} 

\begin{figure*}[!ht]
	\centering
	\includegraphics[width=\columnwidth]{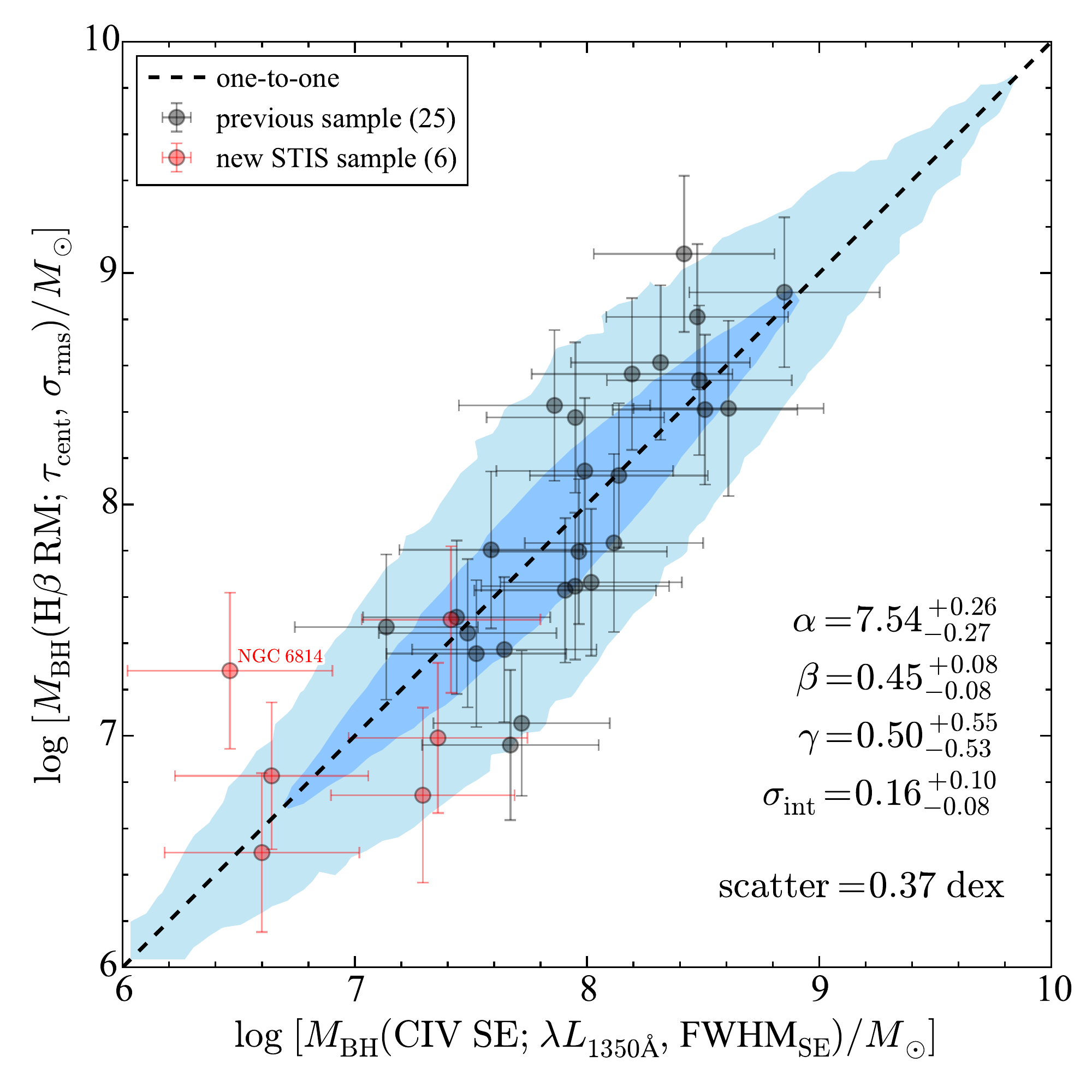} 
	\includegraphics[width=\columnwidth]{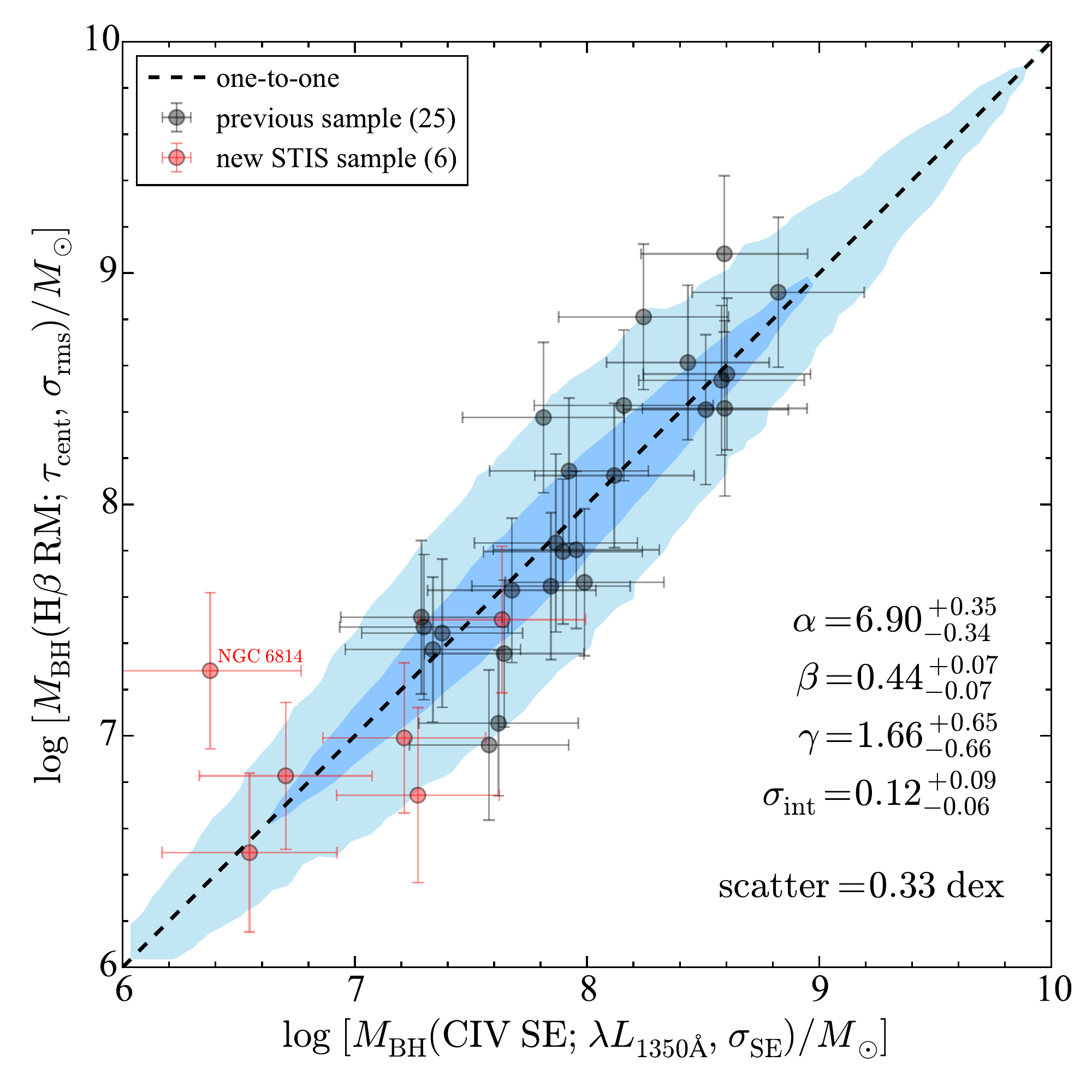} 
	\caption{
		Calibration results of \CIV-based SE BH mass estimators using FWHM (left) and \linedisp\ (right).
		The newly added low-mass AGN sample (6 objects) is indicated with red filled circles while the black filled circles
		represent the previous archival sample of 25 objects from P13.
		Blue shaded contours, whose blue (light blue) area corresponds to the 68\% (95\%) credible region, 
		show posterior predictive distributions under the fitted model as a self-consistency check (see text).
		The resulting regression parameters with the uncertainty estimates and the overall scatter are 
		given in lower right corner in each panel.
		Note that the error bars in the y-axis (RM mass) represent the quadratic sum of 
		the propagated RM measurement uncertainty for the virial product (VP) and the adopted uncertainty for the virial factor,
		and those in the x-axis (SE mass) represent the quadratic sum of 
		the propagated SE calibration uncertainty and the resulting overall scatter.
	}
	\label{fig:calib_FWHM_sigma}
\end{figure*}

\begin{figure*}[!ht]
	\centering
	\includegraphics[width=\columnwidth]{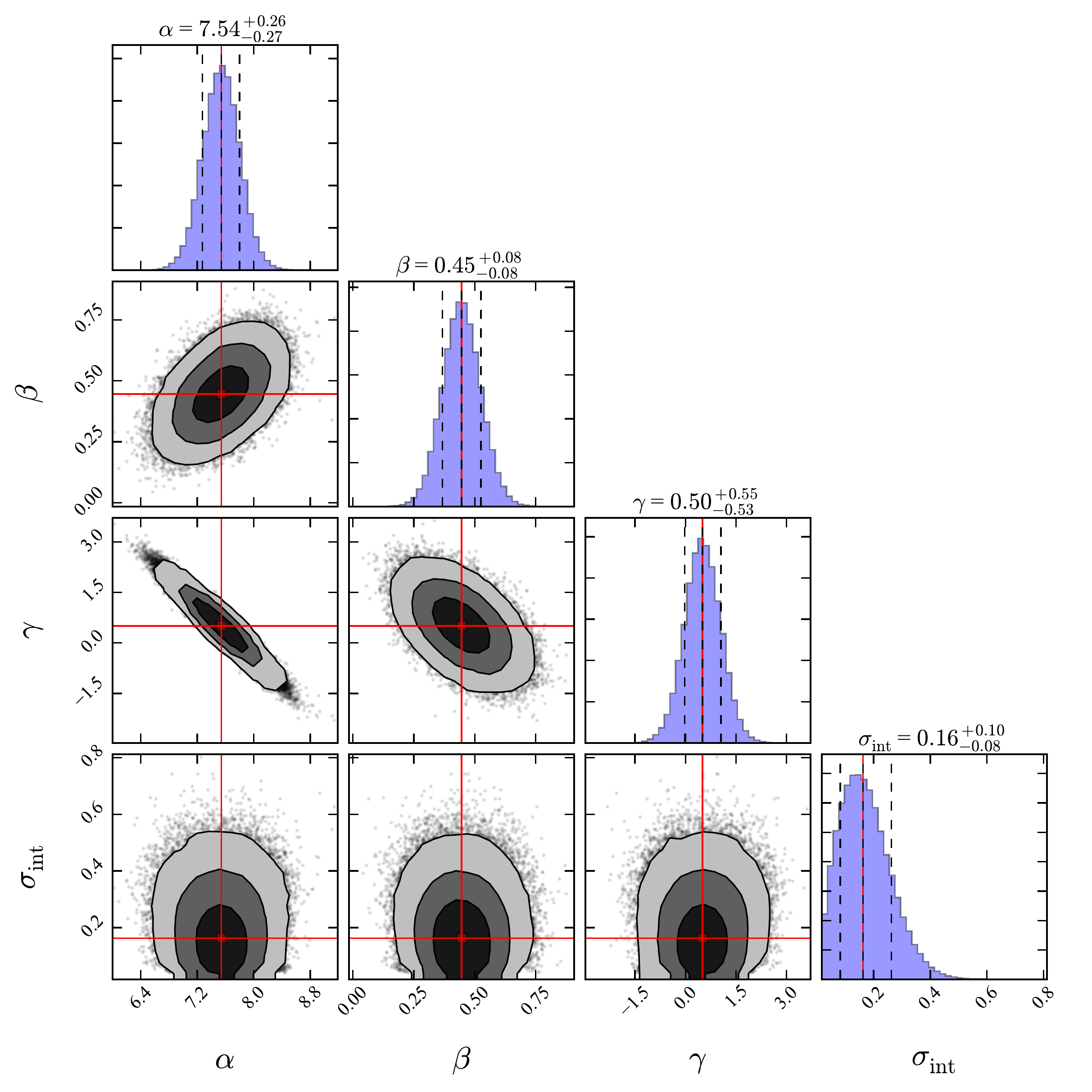} 
	\includegraphics[width=\columnwidth]{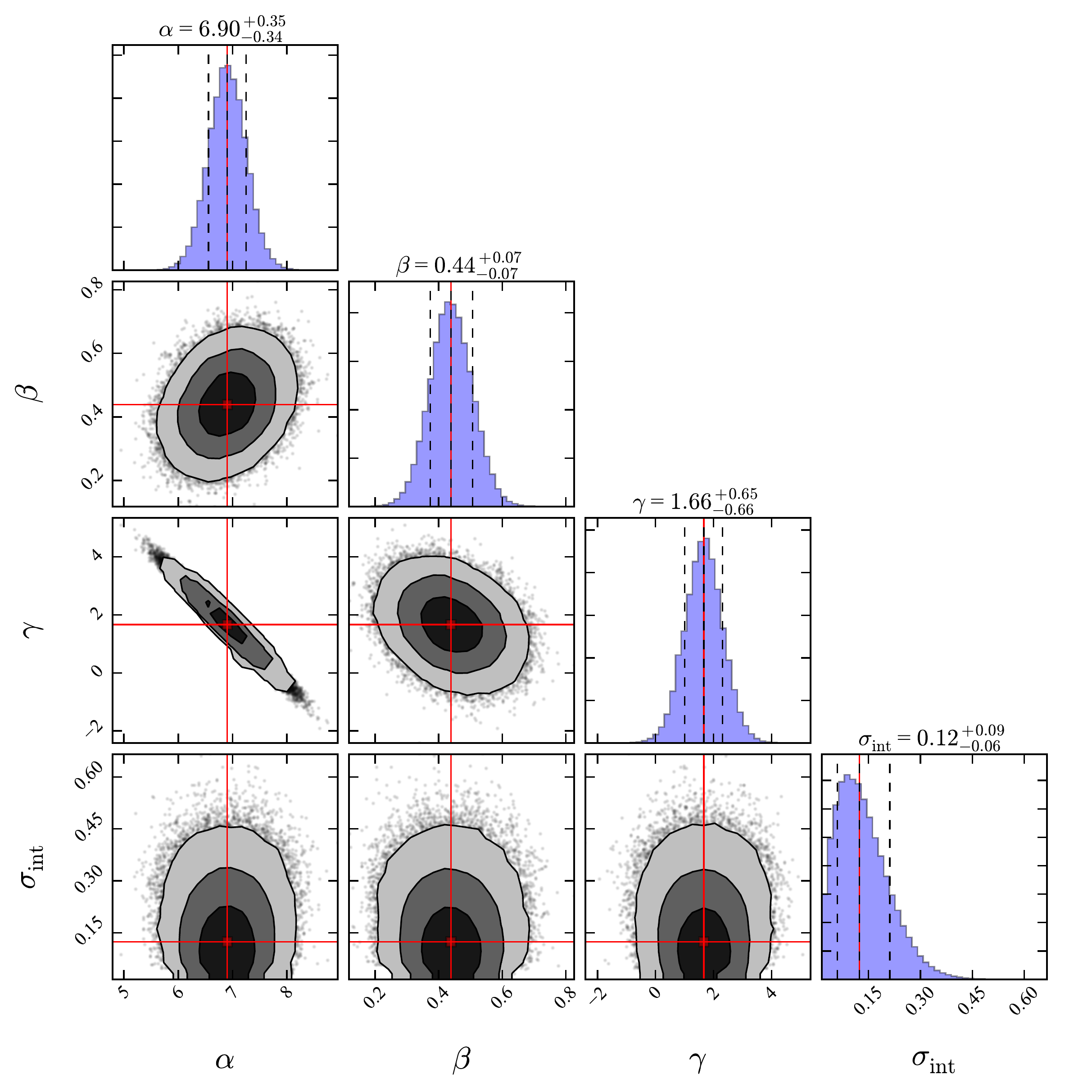} 
	\caption{
		Posterior distributions of the resulting parameters of the calibrations in Fig.~\ref{fig:calib_FWHM_sigma} for the cases of FWHM (left) and \linedisp\ (right).
		The red solid line indicates the posterior median estimate, and the black dashed line marks the uncertainty ranges (i.e., 16\% and 84\% posterior quantiles).
		The 2D marginal distributions (black) of the parameter pairs are shown on the off-diagonal panels,
		while the 1D marginalized histogram (blue) for each parameter from the posterior sample is given on the diagonal panels.
		This figure is made using the \texttt{corner.py}\footnote{\url{https://github.com/dfm/corner.py}}.
	}
	\label{fig:calib_posterior}
\end{figure*}

\autoref{fig:calib_FWHM_sigma} shows the results of calibration of \CIV-based SE BH mass estimators against 
the \Hb-based RM masses using the full sample of 31 local RM AGNs with \autoref{eq:calib},
while \autoref{fig:calib_posterior} presents corresponding marginal projections of each pair of parameters of interest 
with one-dimensional marginalized distributions from the full posterior distribution,
from which parameter covariances are simply identifiable.
We take the best-fit values and uncertainties of parameters of interest ($\alpha,\beta,\gamma,$ and $\sigma_{\rm int}$) from 
posterior median estimates and 68\% posterior credible intervals as recommended by \citet{Kelly2007} and \citet{Hogg+2010}.

In each panel of \autoref{fig:calib_FWHM_sigma}, the two mass
estimates (RM and SE) are fairly consistent. 
The overall scatter of the SE BH masses based on the
calibrated equation using the \CIV\ FWHM (\linedisp) compared to the
RM BH masses are at the level of 0.37 dex (0.33 dex).  This indicates
in general quite good consistency, given the unavoidable
object-to-object scatter of the virial factor $f$ (0.31 dex;
\citealt{Woo+2010}), since we are adopting a single ensemble average
value of the $f$ factor for all objects.

To assess the resulting model fit to the data, 
we present the posterior predictive distribution (blue shaded contour),
which is generated from simulations according to the fit parameters of the model, to check whether 
the posterior prediction reasonably well replicates the original observed data distribution.
As can be seen, the 95\% credible region depicted by the light blue filled contour describes 
most of the data distribution well, except for one outlier, NGC 6814.
We have also calculated posterior predictive $p$-values by following the method of \citet{ChevallardCharlot2016} 
(see also \citealt{BDA3rd} and PyMC User's Guide\footnote{\url{https://pymc-devs.github.io/pymc/modelchecking.html}}).
Using $\chi^2$ deviance as a discrepancy measure,
the Bayesian $p$-value estimates are mostly $\sim0.2$, ranging from 0.14 to 0.26 in this work, 
indicating successful model fits to the data.
Note that there is a problem (e.g., misfit or inadequacy in the descriptive model) 
if the $p$-value is extreme, i.e., $<0.05$ or $>0.95$.
All the calibration results performed in this work for various cases are listed in \autoref{tab:calibration}.
Note that the presence of the single outlier, which is not very extreme,
does not alter any of the conclusions and 
virtually the same result is obtained for give uncertainties if the outlier is removed.

The luminosity slope, $\beta$, is consistent with the photoionization expectation 
(i.e., 0.5) within uncertainties for both FWHM-based and \linedisp-based estimators. 
This may also imply that the size-luminosity relation for the UV 1350 \AA\ continuum has a slope consistent
with the relation for the optical 5100 \AA\ continuum (e.g., $0.533^{+0.035}_{-0.033}$ by \citealt{Bentz+2013}).
The velocity slope, $\gamma$, for the \linedisp-based estimator is close to the virial expectation (i.e., 2) within uncertainties,
while it is not the case for the FWHM-based, whose $\gamma$ value is
consistent with zero, given the uncertainty interval.
This is generally consistent with the calibration result based on \CIV\ FWHM by \citet{ShenLiu2012}, 
who found a much smaller slope (0.242) than 2, although their luminosity dynamic range probed is much higher than ours.
And the \linedisp-based SE masses show overall less intrinsic and total scatter than the FWHM-based masses.
Thus, this indicates that the \CIV\ $\sigma$ is a better tracer of BLR velocity field than the \CIV\ FWHM
because it is closer to the virial relation and shows less scatter in mass estimates.
These results are in overall agreement with \citet{Denney2012}
who found that the \CIV\ FWHM is much more affected by a contaminating non-variable \CIV\ core component 
(see also P13 and \citealt{Denney+2013} for related discussion and interpretation).
The shallower velocity slopes (0.50, 1.66) than the virial expectation (2) would also be expected in part 
due to an additional dispersion of the measured line-of-sight velocities stemming from orientation dependence (see \citealt{ShenHo2014}).

Our final best fits are as follows (see also \autoref{tab:calibration}):
\begin{eqnarray}\label{eq:final_caleq_sigma}
\log \left[\frac{M_{\rm BH} {\rm (SE)}}{M_\odot}\right] 
&~=~& {\bf 6.73}^{+0.07}_{-0.07} 
~+~ {\bf 0.43}^{+0.06}_{-0.06} ~ \log\left(\frac{\lambda L_{1350{\rm \text{\AA}}}}{10^{44}~\rm erg~s^{-1}}\right) \nonumber\\
&~+~& {\bf 2} ~ \log \left[\frac{\sigma (\textrm{\CIV})}{1000~ \rm km~s^{-1}}\right]
\end{eqnarray}
with the overall scatter against RM masses of $0.33$ dex, 
which is defined by standard deviation of mass residuals $\varDelta=\log M_{\rm BH} {\rm (RM)} - \log M_{\rm BH} {\rm (SE)}$,
and
\begin{eqnarray}\label{eq:final_caleq_fwhm}
\log \left[\frac{M_{\rm BH} {\rm (SE)}}{M_\odot}\right] 
&~=~& {\bf 7.54}^{+0.26}_{-0.27} 
~+~ {\bf 0.45}^{+0.08}_{-0.08} ~ \log\left(\frac{\lambda L_{1350{\rm \text{\AA}}}}{10^{44}~\rm erg~s^{-1}}\right) \nonumber\\
&~+~& {\bf 0.50}^{+0.55}_{-0.53} ~ \log \left[\frac{\rm FWHM (\textrm{\CIV})}{1000~ \rm km~s^{-1}}\right]
\end{eqnarray}
with the overall scatter against RM masses of $0.37$ dex.  For the
case of the \linedisp-based estimator, note that the value of $\gamma$
is fixed to be 2 (i.e., consistent with the virial expectation) in our
final analysis because it is consistent with 2 within the uncertainty
estimate when we treat it as a free parameter.  Fixing $\gamma$ to the
physically motivated value helps to avoid possible object-by-object
biases and systematics due to small number statistics, and reduces the
uncertainties of the other resulting regression parameters (see also
P13).  The overall scatter of the \linedisp-based calibration has
virtually unchanged if we fix $\gamma=2$, while that of the FWHM-based
calibration increases considerably if $\gamma$ is fixed to 2 (see
\autoref{tab:calibration}).  

\begin{figure*}[!ht]
	\centering
	\includegraphics[width=\columnwidth]{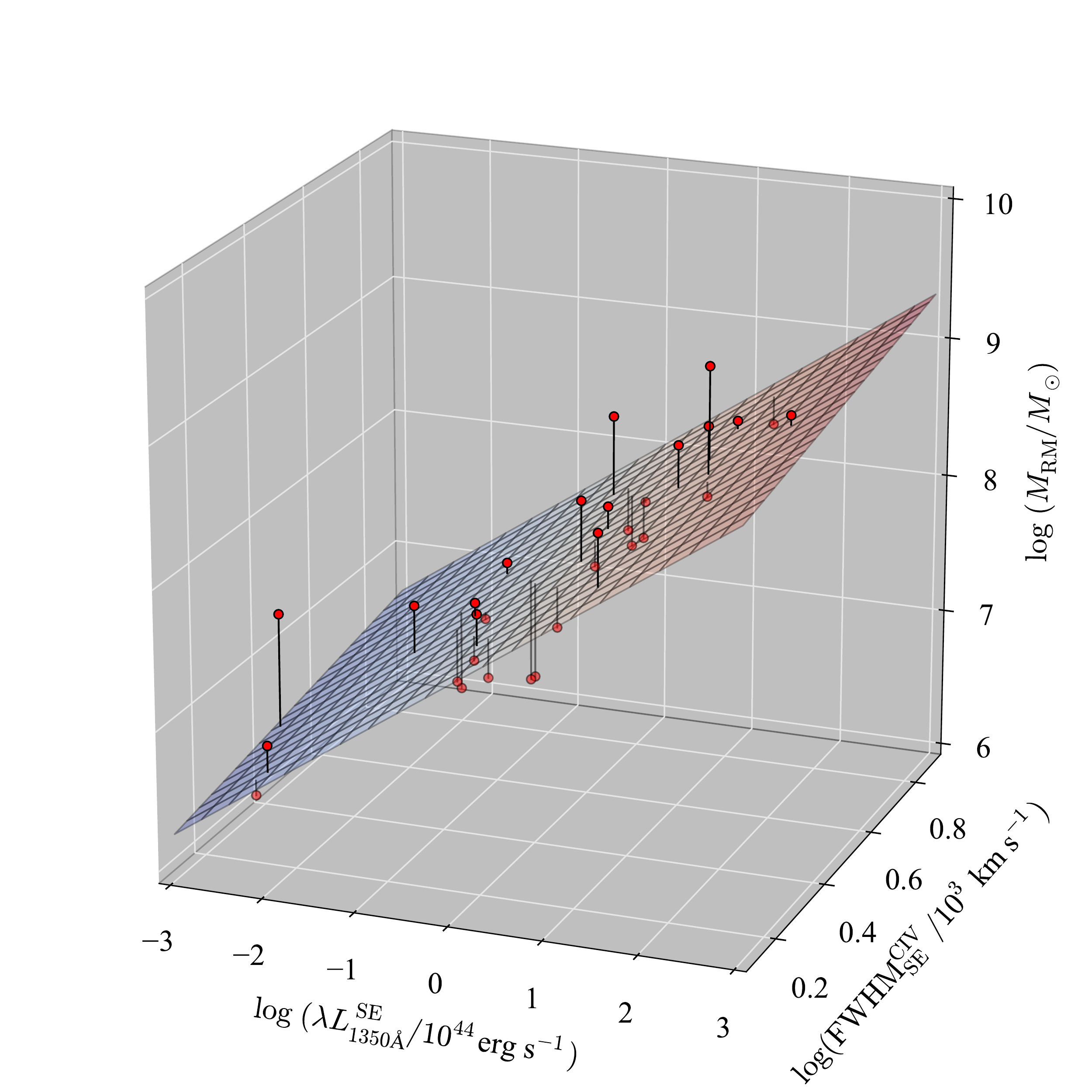} 
	\includegraphics[width=\columnwidth]{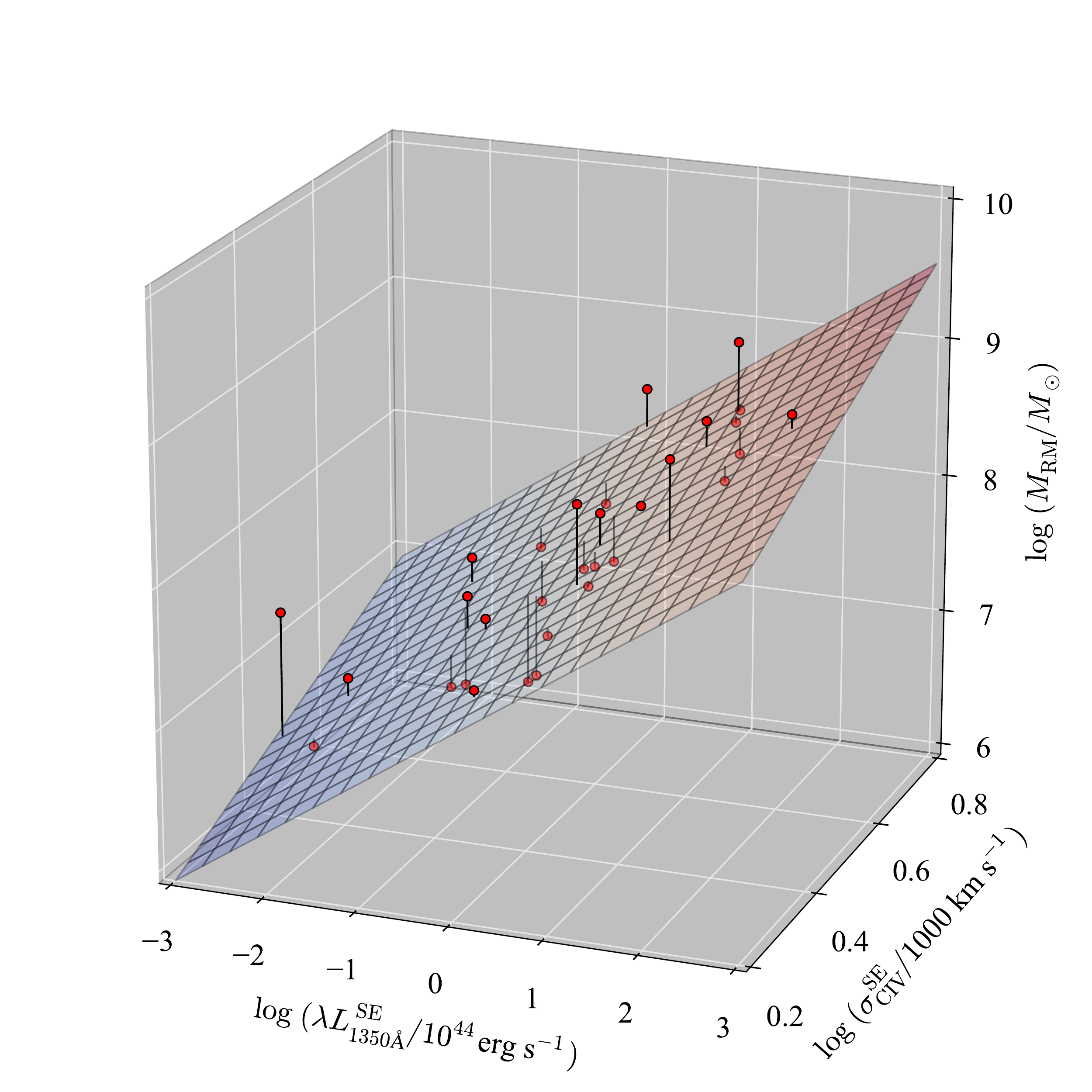} 
	\caption{
		Three-dimensional representation of the calibration results of Fig.~\ref{fig:calib_FWHM_sigma} for clarification.
		The red sphere indicates observed data. The colored tilted plane represents the resulting calibration with \autoref{eq:calib}.
		The black vertical line connecting the data point to the fitted plane shows mass deviation between the observed RM mass and 
		calibrated SE mass.
	}
	\label{fig:calib_FWHM_sigma_3d}
\end{figure*}

\begin{figure*}[!ht]
	\centering
	\includegraphics[width=\columnwidth]{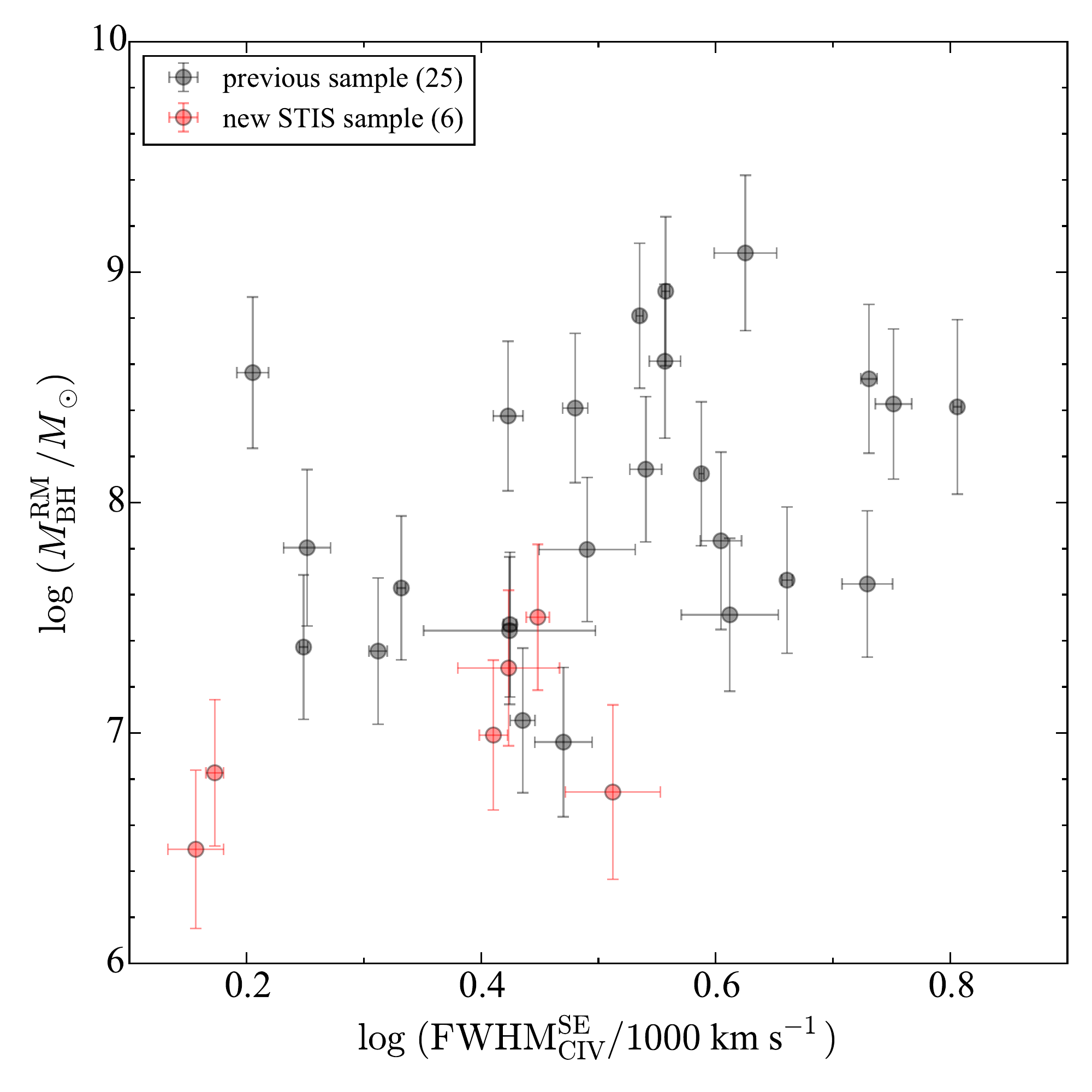} 
	\includegraphics[width=\columnwidth]{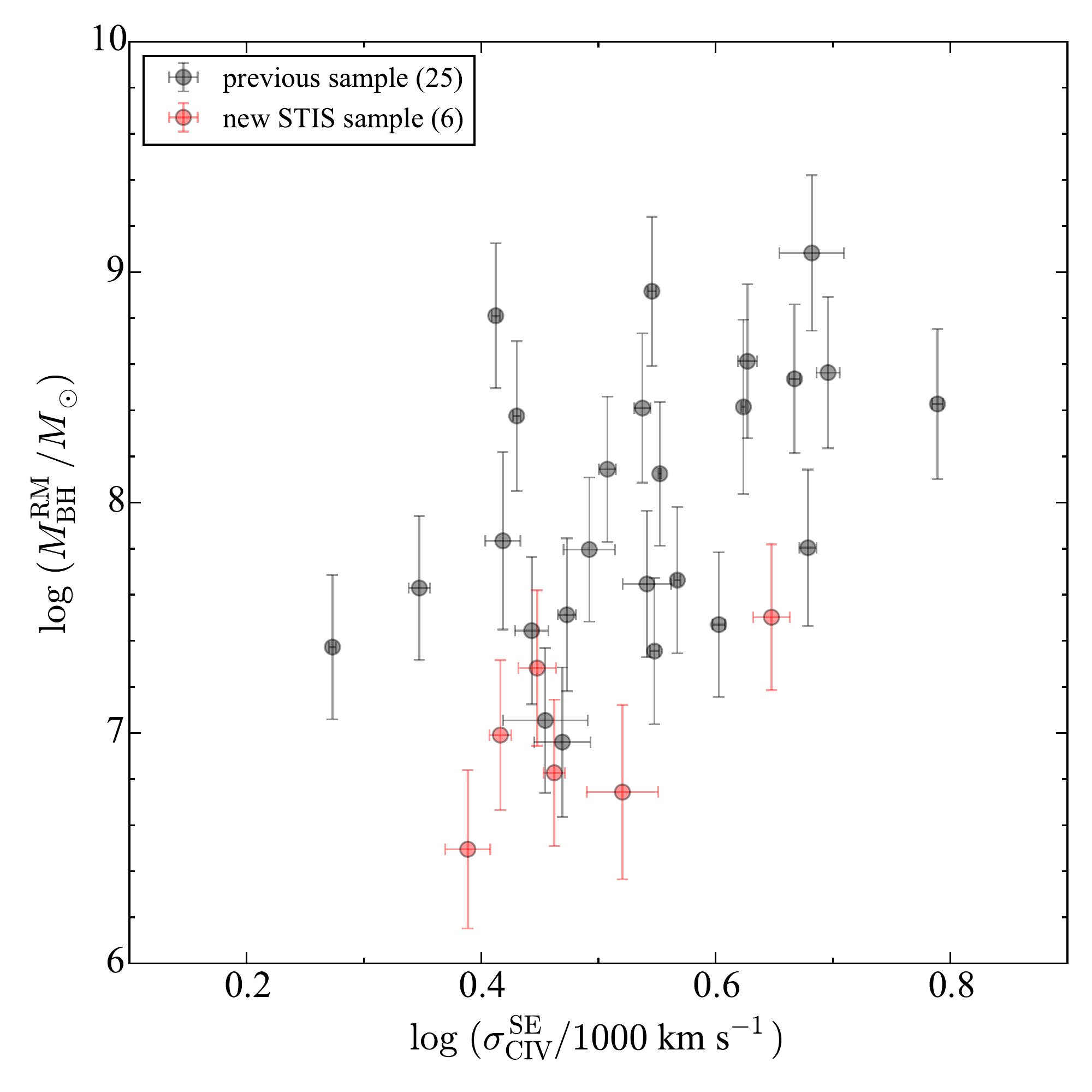} 
	\caption{
		Comparison of SE \CIV\ velocity width measurements, FWHM (left) and \linedisp\ (right), to the observed RM masses.
	}
	\label{fig:comp_M_V}
\end{figure*}

In \autoref{fig:calib_FWHM_sigma_3d}, we show the calibration results
in a three-dimensional space of luminosity, velocity, and mass for clarity.
There is a strong dependence of mass on luminosity,
while there is much weaker dependence of mass on velocity, partly due
to the small dynamic range of broad-line velocity in our sample.
Especially for FWHM, the change of mass as a function of velocity is only marginal.
As expected from the very small value of $\gamma$ for the FWHM-based estimator, 
it seems that \CIV\ FWHM velocity characterization does not
significantly add useful information to mass
estimates, which is consistent with the result by \citet{ShenLiu2012} 
(see also discussions on the importance of line width information by \citealt{Croom2011} and \citealt{Assef+2012}).
It would thus be possible to achieve a comparable level of accuracy in
mass estimates using the luminosity information only, at least for the
present sample.
This much weaker dependence of FWHM on mass than \linedisp\ can also be observed from the projected RM mass-SE velocity plane in \autoref{fig:comp_M_V}, which reinforces that \linedisp\ is a better velocity width measurement for \CIV\ line than FWHM.
Thus, we generally recommend to use the \linedisp-based \CIV\ SE \mbh\ estimator than the FWHM-based
since it is closer to virial relation and shows a better correlation
and less scatter against the RM masses.

In this work (i.e., Equations \ref{eq:final_caleq_sigma}, \ref{eq:final_caleq_fwhm}, and \autoref{tab:calibration}), 
we have used the virial factor, $\log f = 0.71$, taken from \citet{Park+2012ApJS} and \citet{Woo+2010}
for a consistent comparison with the previous result of P13.
If one wants to use a more recent value of the virial factor for BH mass estimates,
e.g., the recently updated virial factor, $\log f = 0.65$, from \citet{Woo+2015},
one can simply subtract the difference ($0.06$) between the adopted virial factors from the normalization, $\alpha$, 
of all the calibration results in this work. Our results can similarly
be rescaled to any other adopted value for virial factor $f$.

\subsection{Comparison with P13}

The final best-fit calibrated equations (\ref{eq:final_caleq_sigma} and \ref{eq:final_caleq_fwhm})
are very similar to those of our previous work (see their equations 2 and 3 of P13).
If we compare the two mass estimates based on both estimators using the same measurements of our sample,
there are very small offsets ($0.02-0.03$ dex) with small scatters of
$\sim0.09$ dex for both FWHM-based and \linedisp-based masses,
which are mostly coming from slight differences in the slopes between P13 and this work.
Although the changes in the adopted virial relation are modest, it is worth noting that this work directly extends the applicability of 
the \CIV-based \mbh\ estimators toward lower BH masses ($\sim 10^{6.5}$ $M_{\sun}$) 
than were present in the P13 sample.

\subsection{Comparison with other linear regression methods}

Advantages of the adopted statistical model using \texttt{Stan} in this work are 
1) using the outlier-robust $t$-distribution as an alternative to the normal distribution for error distributions, and
2) modeling the intrinsic distribution of covariates explicitly with a multivariate $t$-distribution.
To check the performance of our model,
we here provide a comparison of results by performing the same
regression work with other available methods  
(i.e., \texttt{mlinmix\_err} and
\texttt{FITEXY}).

\begin{figure*}[!ht]
	\centering
	\includegraphics[width=\columnwidth]{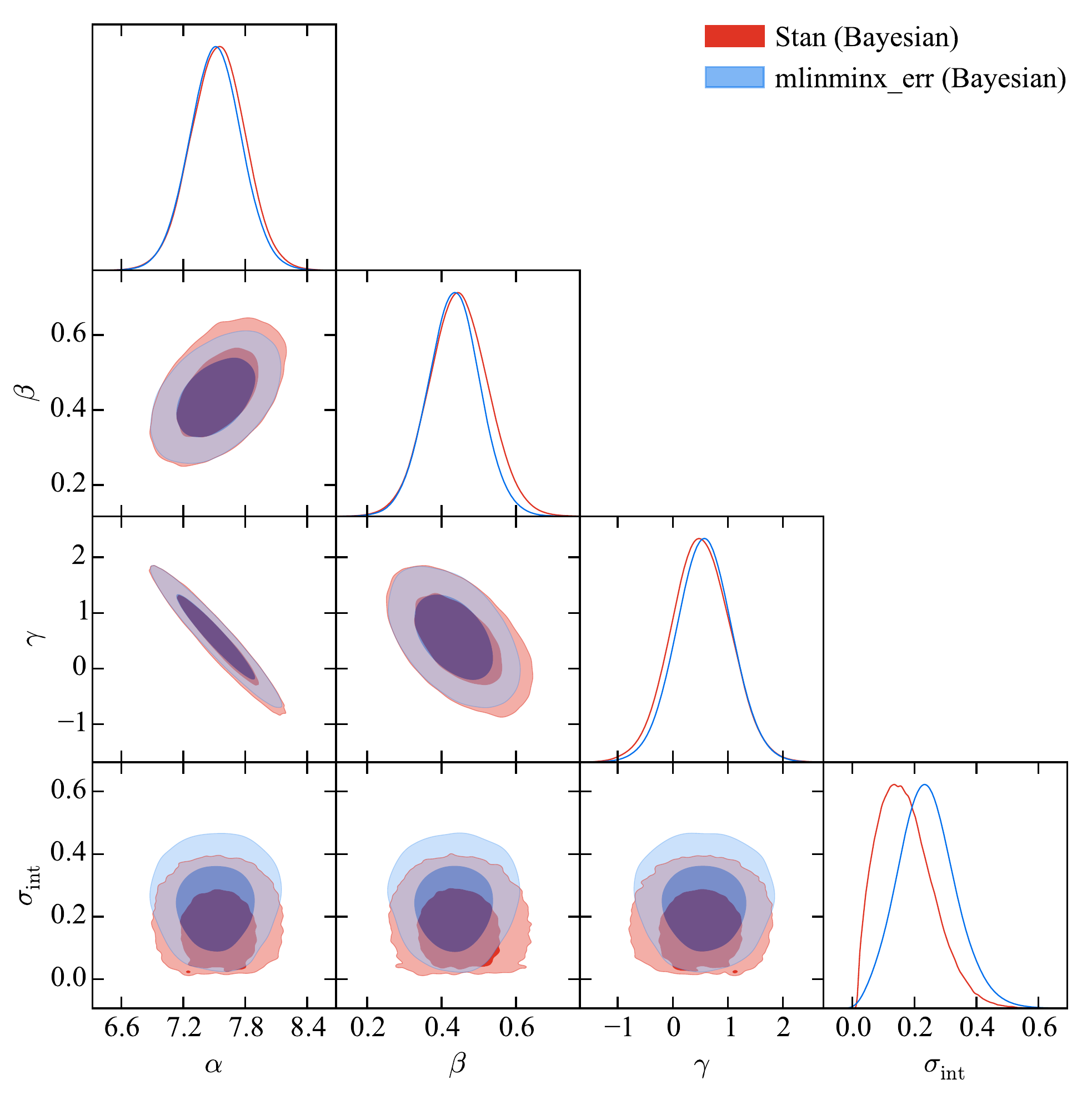} 
	\includegraphics[width=\columnwidth]{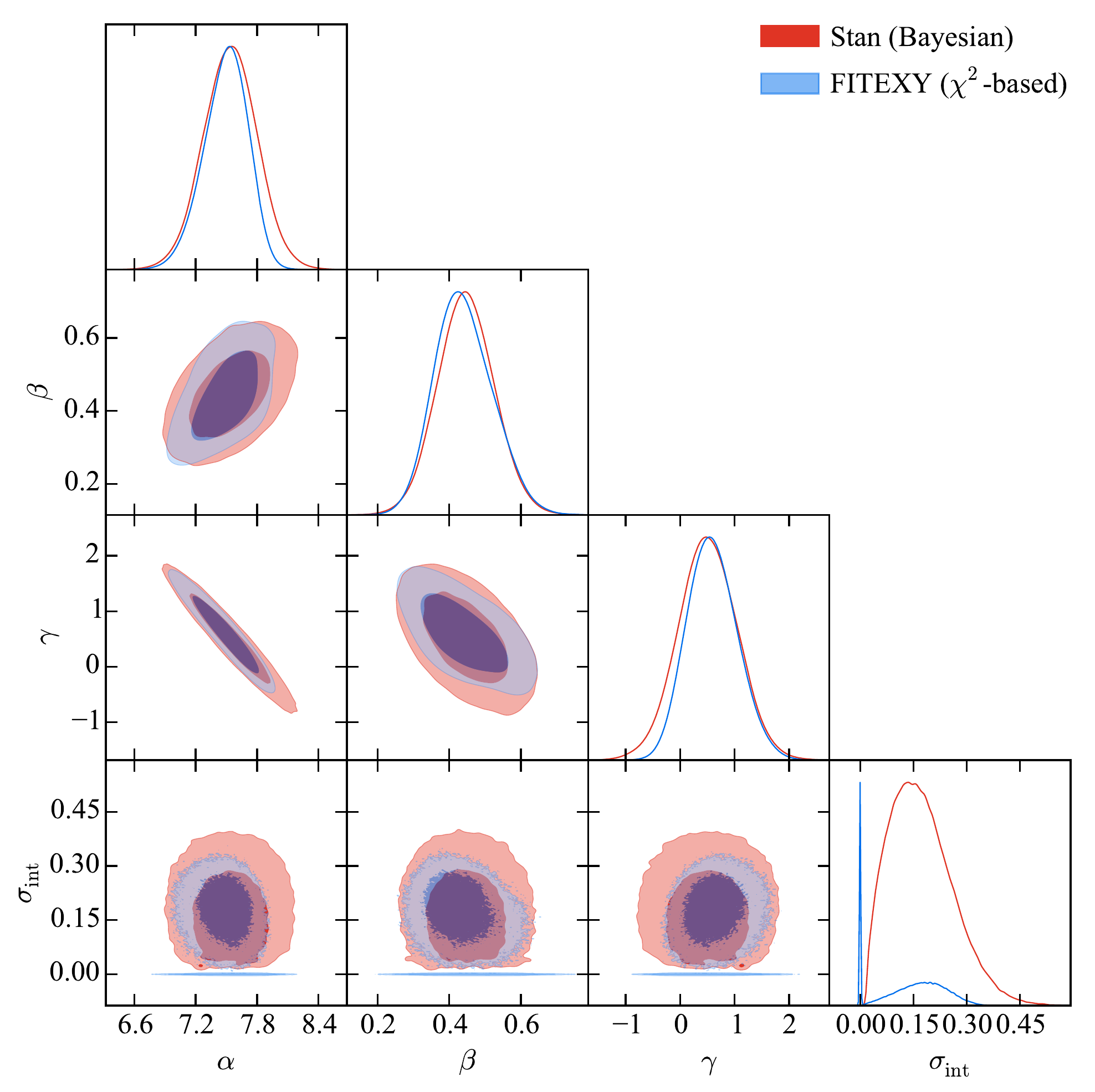} 
	\caption{
		Comparisons of the resulting posterior distributions using \texttt{Stan} to those using 
		\texttt{mlinmix\_err} (left) and \texttt{FITEXY} (right).
		The one- and two-dimensional distributions of a parameter (diagonal panels) and parameter pairs (off-diagonal panels) 
		are shown with the kernel density estimate using the \texttt{GetDist}\footnote{\url{https://github.com/cmbant/getdist}} python package. 
		Note that some amount of smoothing has been applied for a clarity of comparison between the distributions.
	}
	\label{fig:compare_regression_methods}
\end{figure*}

\autoref{fig:compare_regression_methods} compares the resulting posterior distributions of parameters obtained from
three different regression methods for the same data (i.e.,
calibration of the FWHM-based estimator using the 
full sample of 31 local RM AGNs as obtained above).
The \texttt{Stan} Bayesian model implemented for this work (Park 2017 in preparation)
uses the Student-$t$ distributions for measurement errors, intrinsic
scatter, and the covariate distribution model.
The \texttt{mlinmix\_err} method, a Bayesian linear regression code developed by \citet{Kelly2007}, 
employs a normal mixture model for covariate distribution
and assumes Gaussian distributions for measurement errors and intrinsic scatter. 
The \texttt{FITEXY} method, a widely used traditional $\chi^2$-based linear regression method (\citealt{Tremaine+2002}; see also \citealt{Park+2012ApJS} and references therein), also
uses Gaussian distributions for measurement errors and intrinsic
scatter, but has no model specified for the covariate distribution,
and does not take into account possible correlations between measurement errors.

The left panel in Fig.~\ref{fig:compare_regression_methods} shows an overall consistency of the results between the Bayesian methods,
 \texttt{Stan} and \texttt{mlinmix\_err}, except for the distributions of intrinsic scatter.
The quite strong difference of the $\sigma_{\rm int}$ distributions is expected 
because \texttt{Stan} uses $t$-distributed intrinsic scatter
while \texttt{mlinmix\_err} uses normally distributed intrinsic scatter.
By definition, the $\sigma_{\rm int}$ of the $t$-distribution is smaller
than that of the Gaussian distribution 
due to the broader tails of the $t$-distribution (see \citealt{Kelly+2012} and Park 2017 in preparation).
Another noticeable difference between the posterior distributions is
that widths of the probability distributions for regression parameters
obtained from \texttt{Stan} are slightly wider than those from
\texttt{mlinmix\_err}.  Although not significant, this seems to be
indicating more reliable uncertainty estimates with \texttt{Stan},
probably due to the flexibility of the adopted $t$-distributions with
degrees-of-freedom parameters. 
Note that the $t$-distribution widely ranges from the Cauchy distribution to the 
normal distribution as well 
with a varying degrees-of-freedom parameter, but the number of Gaussian components
for the normal mixture model used in \texttt{mlinmix\_err} is fixed to
be a constant (e.g., three by default, although a few Gaussians are
usually well enough to obtain a reasonable description of observed
distributions of many astronomical samples and data, as it is in this
work).

The right panel in Fig.~\ref{fig:compare_regression_methods} also shows a overall consistency between the resulting 
distributions of the Bayesian \texttt{Stan} and the $\chi^2$-based
\texttt{FITEXY} method.
However, underestimates of the parameter uncertainties from \texttt{FITEXY} are a bit more noticeable, 
possibly due to absence of the covariate model description and not accounting for correlations between measurement errors in \texttt{FITEXY} estimates.
The parameter distributions from \texttt{FITEXY} are obtained with a bootstrapping method, so that it may not be a
consistent comparison with the Bayesian posterior distributions.
Many zero values in the $\sigma_{\rm int}$ distribution of \texttt{FITEXY} are also noticeable, which indicates that 
many realizations of bootstrap samples are optimized without the addition of intrinsic scatter.
This behavior is one of the downsides of the $\chi^2$-based
\texttt{FITEXY} estimator, which employs a somewhat ad hoc iterative procedure to determine $\sigma_{\rm int}$
because it cannot be constrained simultaneously with the regression parameters (see \citealt{Kelly2011} and \citealt{Park+2012ApJS}).

The adopted best-fit parameters and uncertainties from the three methods are listed in \autoref{tab:compare_methods} for comparison.
Again, there is no significant difference between the parameter
estimates; 
they are basically consistent with each other within the uncertainties.
The primary reason for this consistency is that
measurement uncertainties for the covariates (line widths and luminosity) are very small in this work
(i.e., only a few percent level on average due to high-quality \HST\ spectra).
Along with it, the resulting covariances between measurement errors are consequently very small as well, 
thus leading to virtually no effect of error correlations on the regression parameter estimates, 
even though there are correlations between measurement errors (see \autoref{tab:UVmeasurement}).
To summarize, 
all three methods (\texttt{Stan}, \texttt{mlinmix\_err}, \texttt{FITEXY}) produce  consistent results in this work
given the small measurement errors of the covariates, 
except for arguably more reliable parameter uncertainty estimates when using \texttt{Stan}.
Although the difference is marginal in this specific work,
more flexible $t$-distributed errors, as well as an explicit covariate model description, 
are generally recommended to get a correct central trend against outliers
by avoiding effects of possible unaccounted systematic errors (see Park 2017 in preparation for details).

\subsection{MAD-based calibration}

Although we prefer line dispersion (\linedisp) to FWHM in measuring \CIV\ line width as investigated above,
one downside of using \linedisp\ is that it requires high S/N data to
accurately fit the line wings.
Noisy data can lead to biases in line width measurements
especially when the line profile has very extended wings as typical of
\CIV\ lines (\citealt{Denney+2013}; see also \citealt{Fine+2010}). 

Recently, \citet{Denney+2016} suggested another way of measuring line
width, the mean absolute deviation (MAD) around the flux-weighted
median wavelength, and suggested it as the most reliable method of
line width measurement for low-quality data.  The MAD is by definition
less affected by core and wing parts of the profile.  Instead, the
middle portions of the velocity profile (relative to the median
velocity) would contribute primarily to determination of line width.
The lower sensitivity of MAD to the line core in comparison with FWHM
is quite useful in order to obtain the least biased line width
measurement when there is a non-varying core component in the
\CIV\ line profile (see \citealt{Denney2012}). Such components are
very hard to identify and remove without using multi-epoch RM data.
Additionally, the MAD has the useful property of being less sensitive
to high-velocity line wings than line dispersion 
(i.e., absolute deviation versus squared deviation as weights). This will be
important when using low-S/N data, which makes accurate
characterization of line wings very difficult.

\begin{figure*}[!ht]
	\centering
	\includegraphics[width=\columnwidth]{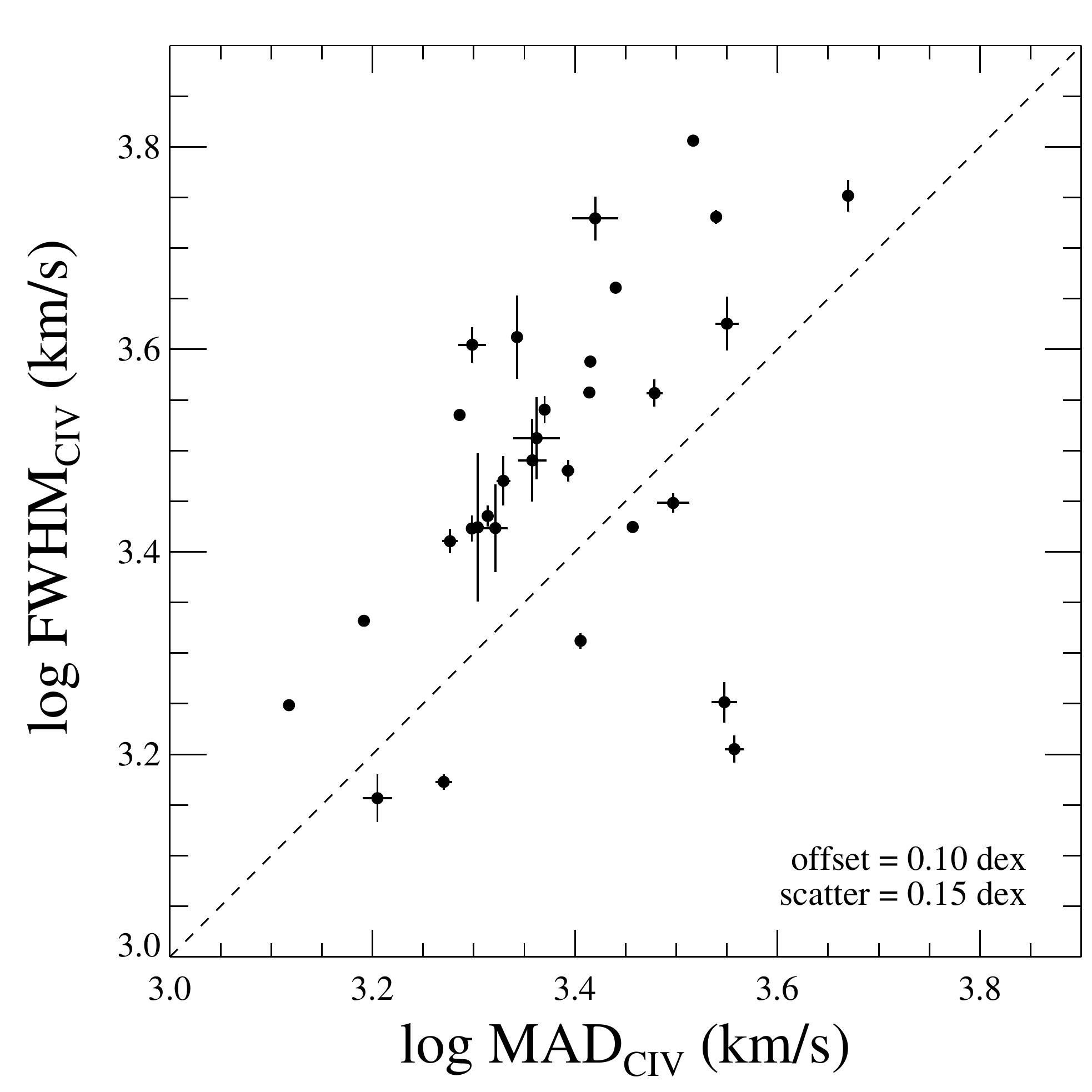} 
	\includegraphics[width=\columnwidth]{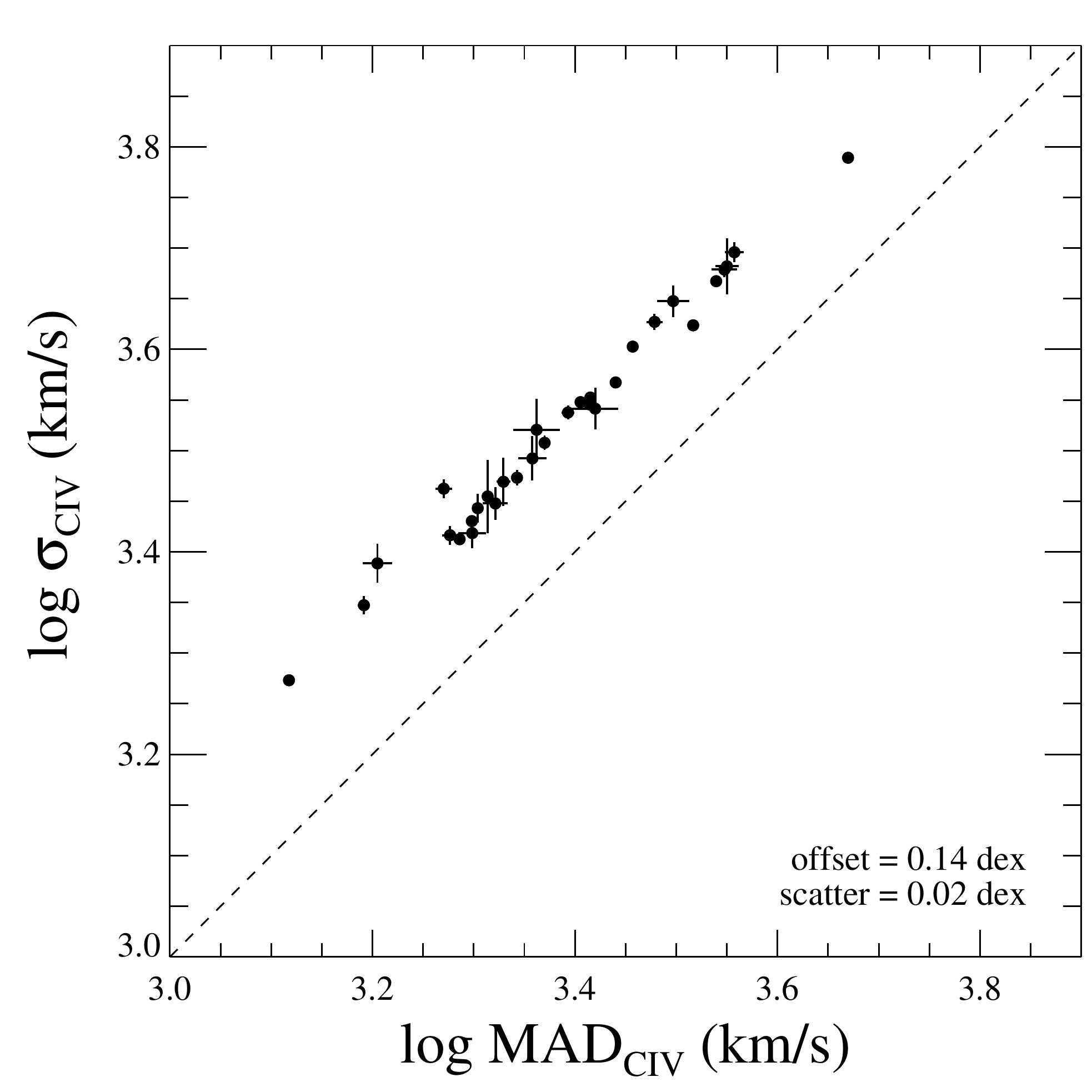} 
	\caption{
		Comparison of MAD to FWHM (left) and line dispersion (\linedisp; right) measurements for our sample of all 31 AGNs.
		The dashed line shows a one-to-one relation.
		The mean offset and 1$\sigma$ scatter are given at the lower right corner.
	}
	\label{fig:mad}
\end{figure*}

Thus, the MAD inherits some of the practical merits of both
\linedisp\ and FWHM, and possibly works better in low-quality data.
We have carried out MAD measurements for the broad
lines in our sample, and we find good consistency between the MAD and
\linedisp\ measurements (\autoref{fig:mad}).  The two measurements are
very nicely correlated with a marginal scatter, while a poor
correlation with a large scatter is observed between the FWHM and MAD.
In this regard, the MAD may be the best line width measurement method
for \CIV\ emission line when using survey-quality spectra as advocated
by \citet{Denney+2016}.  

As it is for the case of \linedisp, the $\gamma$ of MAD is also consistent with $2$ 
within uncertainty if left as a free parameter (see \autoref{tab:calibration}).
Fixing the virial slope to $\gamma=2$, we find the following best-fit
calibration of the SE \CIV\ mass estimator based on MAD as the measure
of \CIV\ linewidth:
\begin{eqnarray}\label{eq:final_caleq_mad}
\log \left[\frac{M_{\rm BH} {\rm (SE)}}{M_\odot}\right] 
&~=~& {\bf 7.01}^{+0.07}_{-0.07} 
~+~ {\bf 0.41}^{+0.06}_{-0.06} ~ \log\left(\frac{\lambda L_{1350{\rm \text{\AA}}}}{10^{44}~\rm erg~s^{-1}}\right) \nonumber\\
&~+~& {\bf 2} ~ \log \left[\frac{\textrm{MAD} (\textrm{\CIV})}{1000~ \rm km~s^{-1}}\right].
\end{eqnarray}
In this case, the overall scatter against RM masses is $0.33$ dex.
The resulting MAD-based calibration and posterior distributions, which are not shown here, are very similar to those of the \linedisp-based,
except for a slight difference in the intercept $\alpha$ (see \autoref{tab:calibration}).

\subsection{Possible biases due to \CIV\ blueshift}

In our calibration sample (i.e., local RM AGNs), \CIV\ blueshifts are basically insignificant (see \citealt{Richards+2011,Shen2013}),
so that our calibration based on the local RM AGNs is relatively free of possible biases stemming from 
the effect of large blueshift.
However, the applicability of this calibration to high-z quasars may
be uncertain
because large \CIV\ blueshifts are known to be common in high-z, high-luminosity quasars (see, e.g., \citealt{Richards+2002}).
Available \CIV-based \mbh\ estimators have been used for measuring BH masses of a statistical sample of such high-z AGNs,
simply based on assumption and extrapolation without a direct test.
The best way of investigating and possibly correcting for the effect of \CIV\ blueshifts on BH mass estimates
would be using direct \CIV\ RM data (see \citealt{Denney2012} for the case of local AGNs).
The number of AGNs having direct \CIV\ RM observations is, however,
very limited, due to the major practical difficulties of obtaining RM
measurements for high-$z$, high-luminosity AGNs as well as the
difficulty of obtaining space-based UV monitoring data for low-$z$ AGNs.

Instead, recently, \citet[][see also \citealt{ShenLiu2012}]{Coatman+2016,Coatman+2017} have provided 
a new empirical correction to \CIV\ FWHM-based BH mass estimators as a function of \CIV\ blueshift
by comparing SE \CIV\ measurements to SE \Ha\ measurements.

\begin{figure*}[!ht]
	\centering
	\includegraphics[width=\columnwidth]{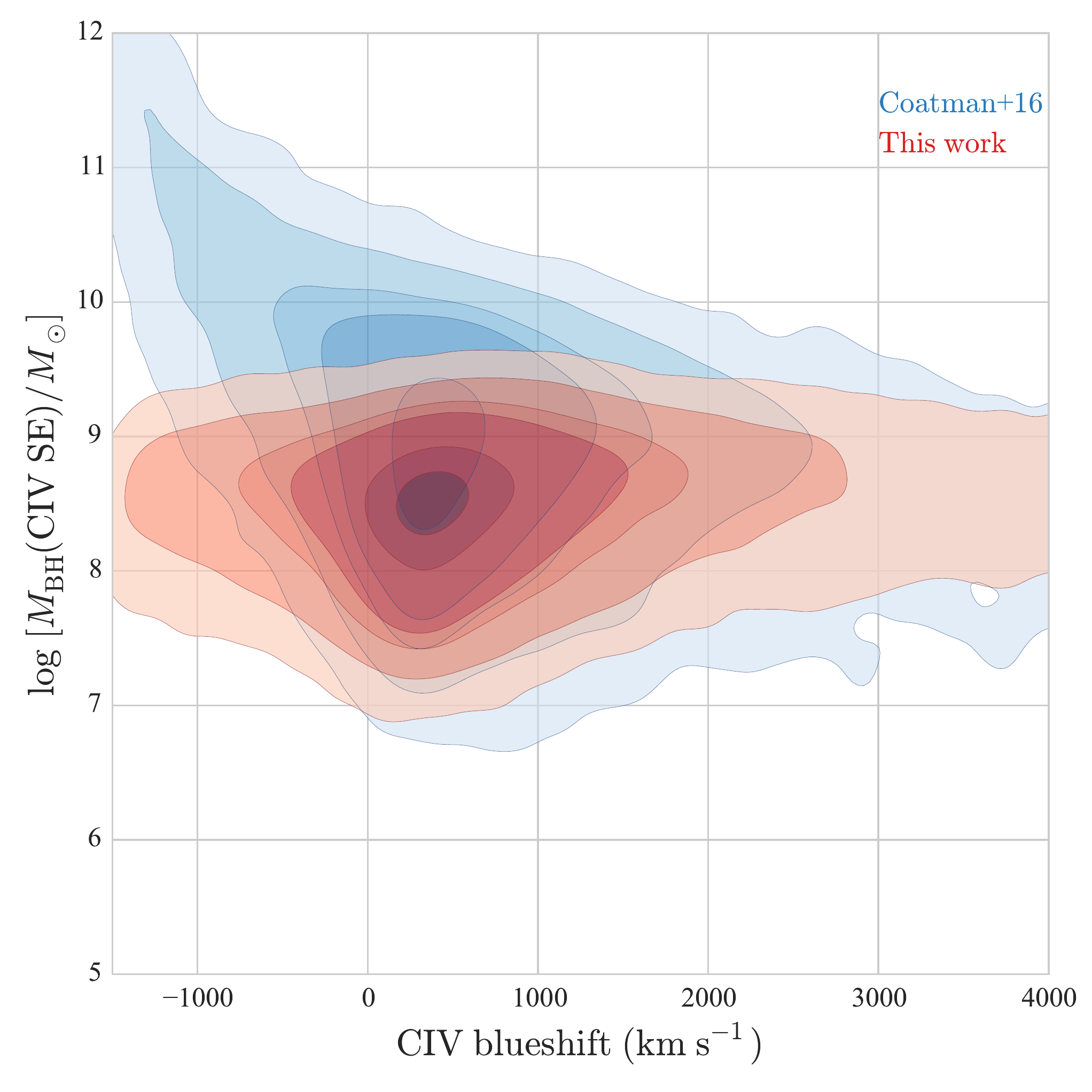} 
	\includegraphics[width=\columnwidth]{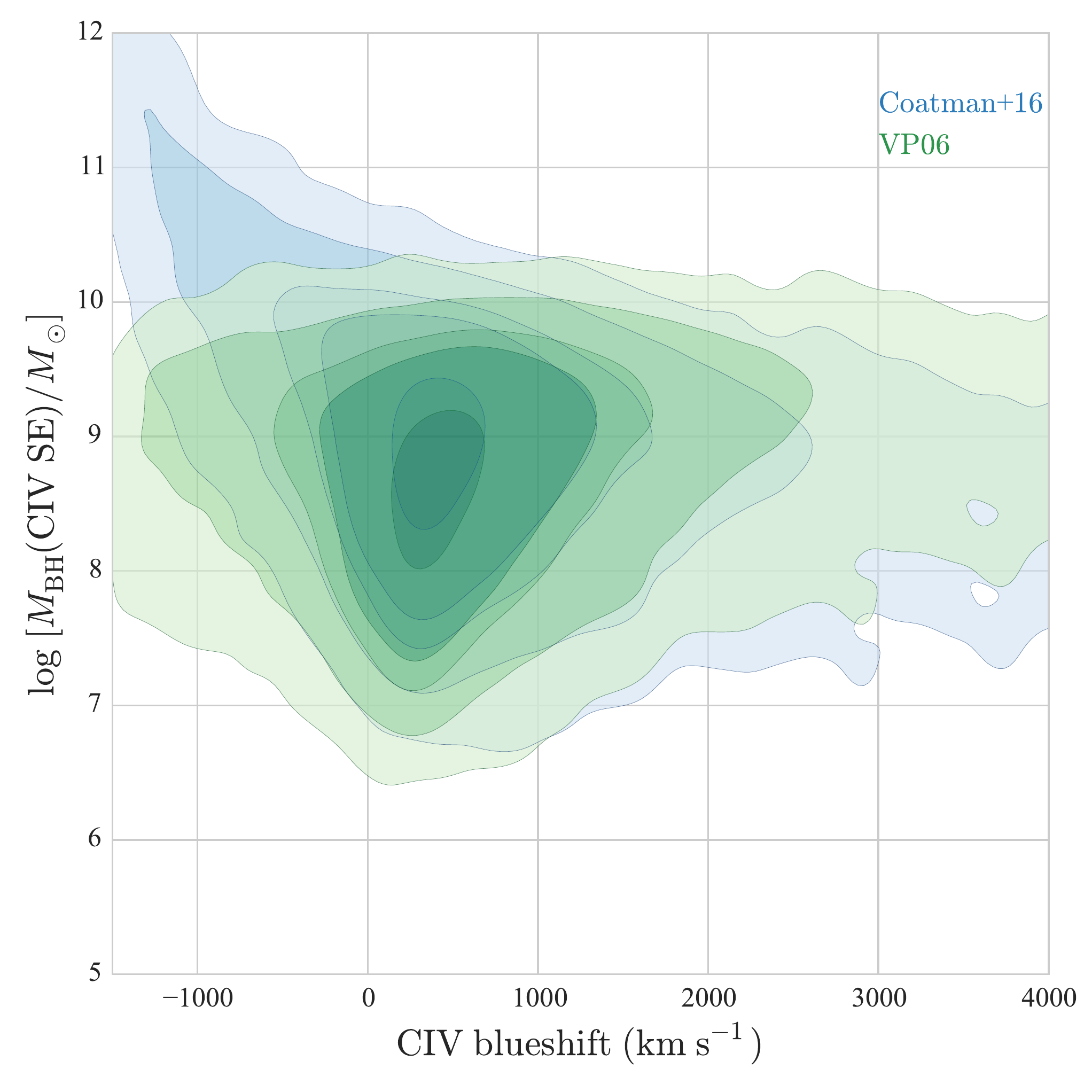} 
	\caption{
		Distributions of \CIV\ FWHM-based \mbh\ estimates as a function of \CIV\ blueshift 
		for SDSS quasars from the DR9 BOSS quasar catalog. 
		Most of them are in the redshift range $2\lesssim z \lesssim3$.
		The left panel compares BH masses computed from the new calibration of this work (red) to those using the 
		blueshift-corrected recipe from \citet{Coatman+2017} (blue) 
		while BH masses using the calibration of VP06 (green) are compared in the right panel.
		This figure is made using the \texttt{Seaborn}\footnote{\url{http://seaborn.pydata.org/}} python package.
	}
	\label{fig:MBHdist}
\end{figure*}

In \autoref{fig:MBHdist}, we compare the overall distributions of \CIV\ FWHM-based BH mass estimates 
as a function of \CIV\ blueshift using the spectral measurements of DR9 BOSS quasars\footnote{provided by Yue Shen at \url{http://quasar.astro.illinois.edu/BH_mass/dr9.htm}}. 
The blue shaded contour presents BH masses computed from the blueshift-corrected formula from \citet{Coatman+2017},
while the red and green shaded contours show those calculated with the new updated recipe in this work and the original VP06 equation, respectively.
As can be seen, at large blueshift ($\gtrsim2000$ \kms), 
our estimator, which does not take into account \CIV\ blueshift, 
produces a similar mass distribution to the blueshift-corrected distribution.
Note that overestimated BH masses from the VP06 estimator are reduced by correcting for the blueshift effect on \CIV\ FWHM  \citep{Coatman+2017}.
Thus, our locally calibrated FWHM-based \mbh\ estimator would be applicable to a sample of high-$z$, high-luminosity quasars 
having high \CIV\ blueshifts (e.g., $\gtrsim2000$ \kms), giving a consistent mass scale on average.

At the range of small blueshfit ($\sim 0-1000$ \kms), where our calibration sample is distributed,
the overall mass scale from our estimator is smaller than those of
VP06 and \citet{Coatman+2017} in the high mass regime
($\gtrsim10^{8.5}M/M_{\sun}$) and larger in the low mass regime ($\lesssim10^{8.5}M/M_{\sun}$).
This trend has been described in detail by P13.
Arguably, our calibration in this work (and P13) has resulted in a
better agreement (overall mass scale) with RM masses than that of VP06
in terms of intrinsic scatter and using the higher quality dataset of
the updated sample, at least in the mass range
($10^{6.5-9.1}M/M_{\sun}$) where the calibration has been performed.
Note that neither our calibration nor that of VP06 (also
\citealt{Coatman+2017}, which is based on VP06 calibration) is
directly confirmed in the very high-mass BH regime
($\gtrsim10^{9}M/M_{\sun}$ where most of the Coatman sample is) due to
a lack of high-mass RM AGNs in the calibration samples.  All these
trends are also shown in Fig. 13 of \citet{Coatman+2017}.

It is also worth noting that 
striking mass increase toward negative blueshift from using the equation of \citet{Coatman+2017} is obviously unreliable, 
as already discussed by \citet{Coatman+2017}, due to their insufficient dynamic range of blueshift.
This \CIV\ blueshift-corrected recipe should therefore not be used for  objects with negative \CIV\ blueshift.

As shown by \citet{Coatman+2016}, the \Ha\ line seems to be also systematically changing as a function of \CIV\ blueshift,
although the \Ha\ line measurements may not be very accurate due to very low S/N for the \Ha\ spectral region
(mostly $\lesssim10$ per resolution element; see their Table 1).
If this is true, calibrating SE \CIV\ line to SE \Ha\ (or \Hb) line as
done by \citet{Coatman+2016} would be flawed.
In other words, correcting \CIV\ FWHM as a function of blueshift against \Ha\ FWHM would still be biased 
since the \Ha\ FWHM is also correlated with \CIV\ blueshift.
As an ultimate goal, calibrating SE \CIV\ mass estimators against
direct \CIV\ RM data (or indirectly against RM Balmer line if
\CIV\ RM data is unavailable) for a much larger sample including
high-luminosity, high-\CIV\ blueshift AGNs will be the best way to
improve the SE mass method.
However, given the difficulty of obtaining many direct \CIV\ RM
measurements for both low- and high-$z$ AGNs and determining accurate
blueshifts (and systemic redshifts) (see, e.g.,
\citealt{Denney+2016z,Shen+2016velshift}), our simple calibration of
SE \CIV-based \mbh\ estimators will still be useful when estimating BH
masses from \CIV\ observations of AGNs over a wide range of redshift
and luminosity.

\subsection{Comparison to other prescriptions}
\begin{figure*}[!ht]
	\centering
	\includegraphics[width=0.85\columnwidth]{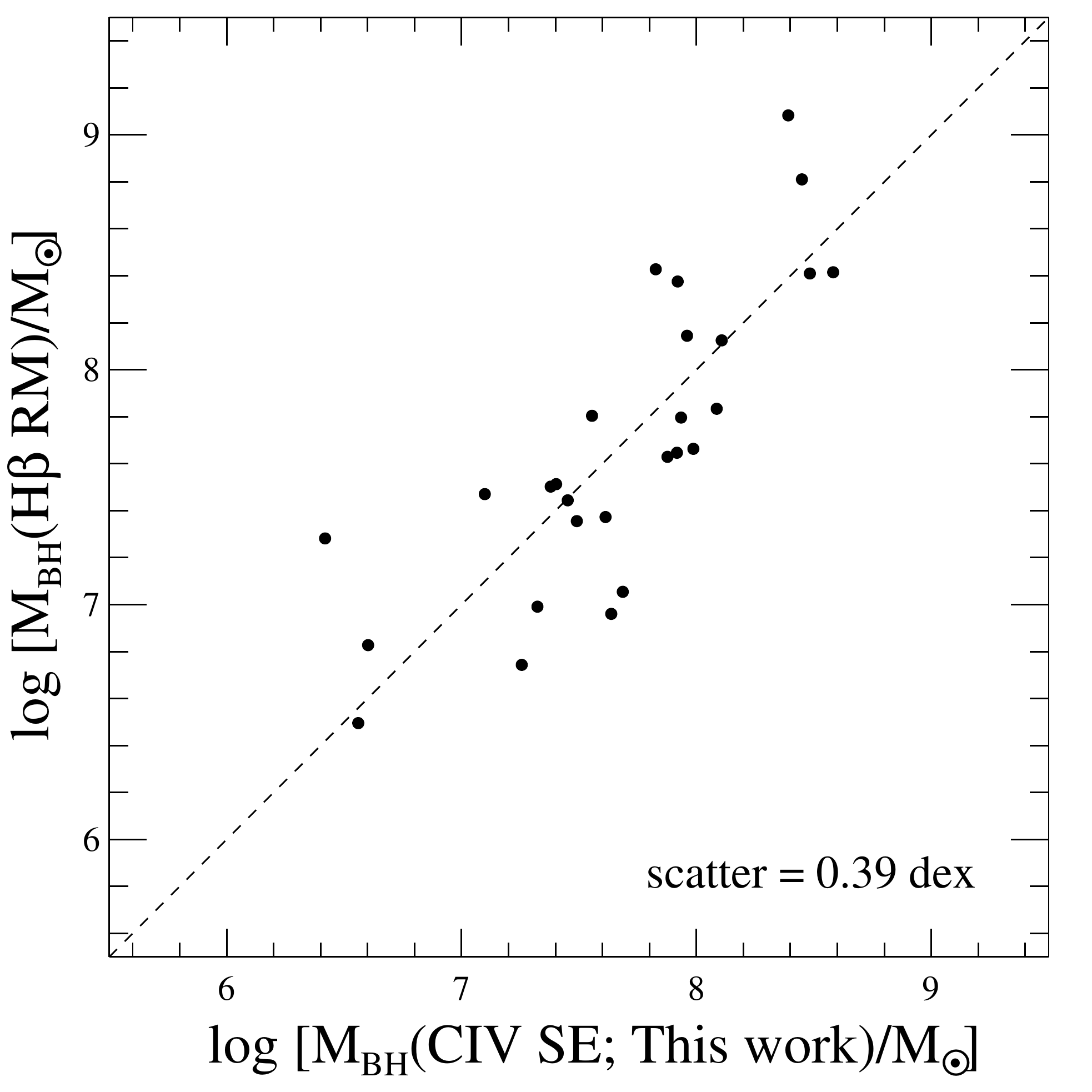}\\
	\includegraphics[width=0.85\columnwidth]{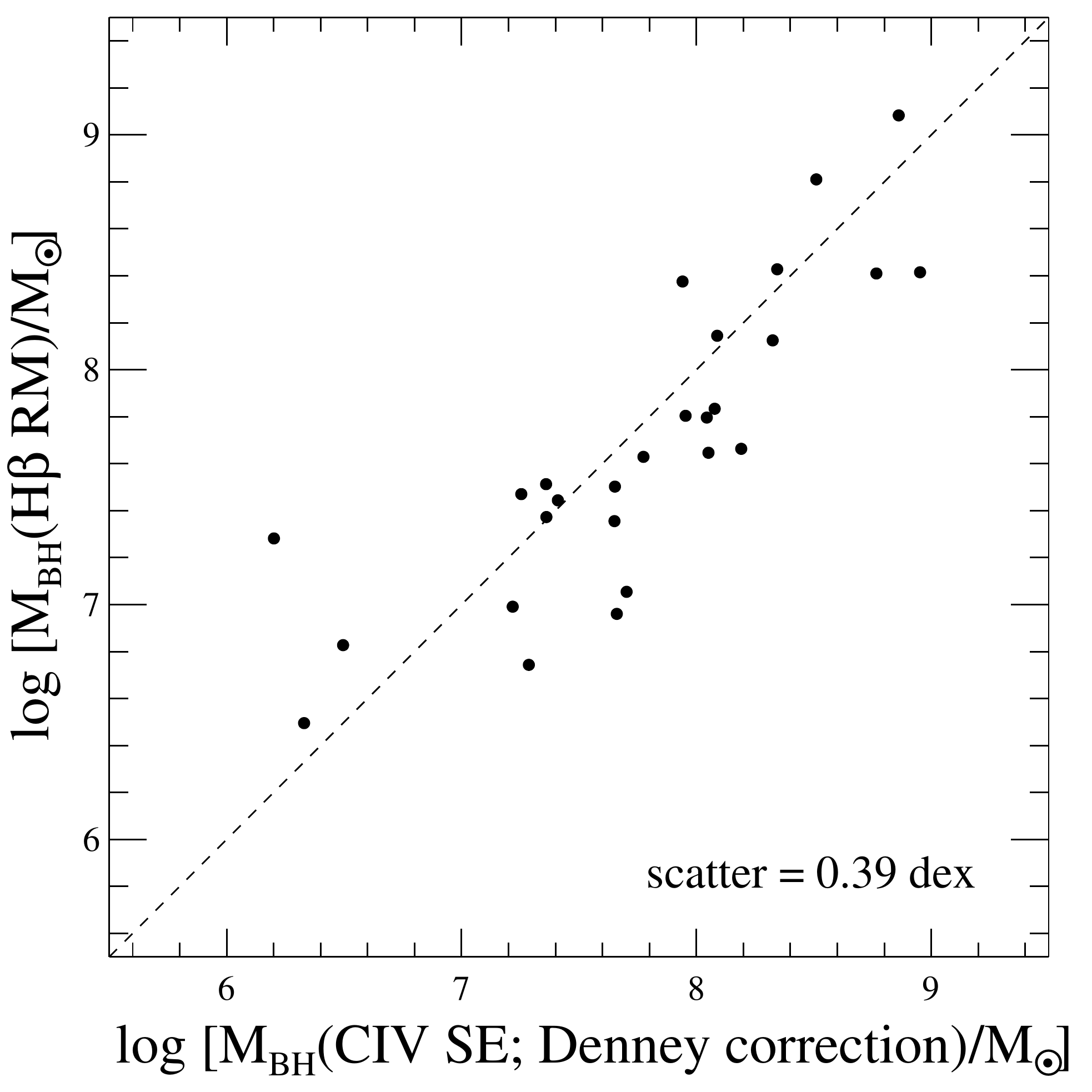}\\
	\includegraphics[width=0.85\columnwidth]{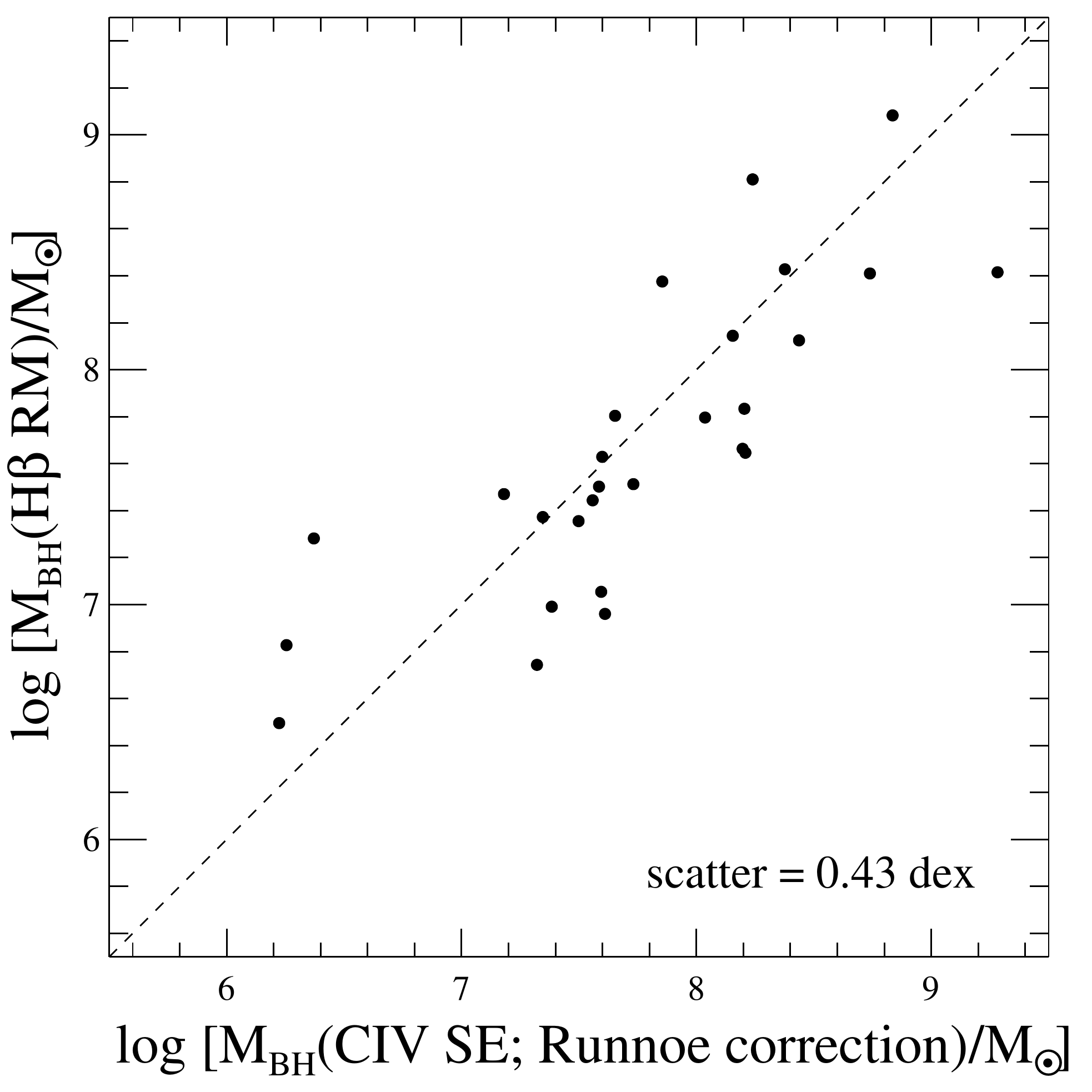} 
	\caption{
		Comparisons of \CIV\ FWHM-based SE \mbh\ estimates using our new calibration (top),
		the \CIV\ line shape based correction by \citet{Denney2012} (middle), 
		and the $\lambda1400$ feature based correction by \citet{Runnoe+2013} (bottom) 
		to the \Hb\ RM-based BH masses for our sample of local RM AGNs. 
		The dashed line shows a one-to-one relation, and the 1$\sigma$ scatter is 
		given at the lower right corner in each panel.
	}
	\label{fig:compareMBHcorrections}
\end{figure*}

In \autoref{fig:compareMBHcorrections}, 
we compare the \Hb-RM based BH masses to the \CIV\ FWHM-based SE BH masses 
from the corrected prescriptions presented by \citet[][their Equation 1]{Denney2012} 
and \citet[][their Equation 3]{Runnoe+2013}.
Note that we here use our sample of the local RM AGNs, except for four objects 
(PG 0026+129, PG 0052+251, PG 1226+023, PG 1307+085) that have 
not enough spectral coverage to measure the $\lambda1400$ feature (see Figure 1 of P13).
The peak flux of the emission line blend of \ion{Si}{4}$+$\ion{O}{4}] (i.e., $\lambda1400$ feature)
has been measured by fitting it with a local power-law continuum and multi-Gaussian functions 
following the same method by \citet[][see also \citealt{Shang+2007}]{Runnoe+2013}.
As a direct comparison, we also show the \CIV\ FWHM-based masses 
using our new calibration (\autoref{eq:final_caleq_fwhm}).

The SE BH masses using the \CIV\ line shape (FWHM/\linedisp) based correction by \citet{Denney2012} shows 
the overall scatter of $0.39$ dex, which is the same as that of our calibration.
However, this corrected prescription is not practically useful because it requires a \linedisp\ measurement, 
as well as FWHM, to obtain the shape measurement for the correction to \CIV\ masses.
One can use \linedisp\ directly, if it is available, rather than using FWHM.
The slightly larger scatter of $0.43$ dex is observed for the case of the peak flux 
ratio ($\lambda1400$/\CIV) based correction by \citet{Runnoe+2013}.
The effect of the correction is less pronounced for our sample, which is not surprising
from the investigation by \citet{Brotherton+2015}, who found that the peak flux ratios 
measured for the RM AGN sample did not correlated with the difference between 
\Hb\ and \CIV\ velocity widths.
Our simple calibration is again a useful practical tool in the situation that such 
additional measurements are not available.

\section{SUMMARY AND DISCUSSION}\label{sec:summary}

We have updated the calibration of \CIV-based SE \mbh\ estimators
based on an enlarged AGN sample 
with high-quality \HST\ UV spectra and using Bayesian linear regression analysis.
As an extension of the previous work of P13, 
there are several improvements over the previous calibration:
the sample now covers masses down to $\sim 10^{6.5}M_{\sun}$ with
measurements from high-quality and quasi-simultaneous UV-to-optical STIS spectra, and
we have used a  Bayesian linear regression method 
to perform outlier-robust inference and take into account covariate distributions and 
possible correlations between measurement errors.

The results presented in this work are consistent with our previous
(P13) and are also in line with \citet{Denney2012} and
\citet{Denney+2013}.
We generally recommend use of the \linedisp-based or MAD-based
\CIV\ \mbh\ estimators, when the measurement are available, since they
are better proxies for BLR velocity field (close to the virial
relation) and show less scatter in mass estimates than the FWHM-based
measurements.  Using \linedisp\ or MAD rather than FWHM for \CIV\ line
width measurement is supported by the fact that accurately decomposing
and removing a \CIV\ narrow component, if any, is difficult to
accomplish with single-epoch spectra.  Thus, to avoid possible biases
due to a possible \CIV\ core component \citep{Denney2012}, using the
line width measurement that is least affected by uncertain line core
(i.e., \linedisp\ or MAD) appears to be the best approach at present.
Measuring \linedisp\ requires high-quality data to accurately
characterize line wings, while MAD is less sensitive to high-velocity
wings.  \CIV-based SE \mbh\ estimators are commonly applied to survey
quality data (e.g., SDSS quasars), where \linedisp\ might not be
robustly measured.  FWHM is relatively straightforward to measure even
in low S/N data, and FWHM measurements are usually provided as the
primary measure of line width in survey catalogs (see, e.g.,
\citealt{Shen+2011,Paris+2017}).  However, this does not mean that
\CIV\ FWHM provides an unbiased estimate of the \CIV\ virial velocity
in low-quality data.  Furthermore, \citet{Denney+2013} showed that
high-quality data do not improve \CIV\ FWHM-based BH mass estimates,
and the best-quality \CIV-based BH masses are obtained using 
\linedisp\ values measured from high-quality data.

All the calibrations presented in this work and other similar works from literature are, however,
subject to sample biases from incompleteness of the calibration samples. 
Although we were able to expand the BH mass  range to lower masses
compared with previous work,
there is still a lack of calibration objects at very high masses
($\gtrsim10^9M_{\odot}$) and in the regime of strong blueshifts.
It is thus important to conduct direct tests of the reliability of
extrapolation of this calibration toward high-redshift,
high-luminosity quasars, which commonly have high BH mass and/or
strong \CIV\ blueshift, as discussed by \citet{Richards+2011} and
\citet{Shen2013}.  Current and future multi-object RM programs including SDSS-RM
\citep{Shen+2015SDSSRMoverview}, OzDES \citep{King+2015}, and MSE
\citep{McConnachie+2016} will help to improve this situation by
providing direct reverberation measurements for rest-frame UV lines in
large numbers of quasars.
There have been other efforts to improve the calibration of the
\CIV\ SE mass scale by taking into account \CIV\ blueshifts (e.g.,
\citealt{ShenLiu2012,Coatman+2016,Coatman+2017}) and by making
use of other measured quantities including UV-to-optical color, line
shape, and nearby line peak flux ratio in the calibration of the SE
method (\citealt{Assef+2011, Denney2012, Runnoe+2013,
  Brotherton+2015}).
However, the most fundamental and best way of
achieving an accurate calibration of \CIV-based BH mass estimators will
only be from direct \CIV\ reverberation mapping of significant samples
of AGN.

A spectrum with broad wavelength coverage observed simultaneously is
essential in order to 
accurately investigate the rest-frame UV-to-optical continuum and emission lines and velocity offsets between them
without suffering from systematics due to intrinsic AGN variability 
(see, e.g., \citealt{Ho+2012,Capellupo+2015,Mejia-Restrepo+2016} for such datasets).
However, even with such data at hand, it is difficult to achieve good
continuum fits over the region from $\sim3100$ \AA\ to $\sim4000$
\AA\ if fitting the entire spectral region at once, due to the
incompleteness and limitation of the currently used AGN
\FeII\ emission templates.  No available template covers the full
UV/optical range, which is essential to constrain the Balmer continuum
and \FeII\ emission accurately and continuously. There is a need for
further improvement in \FeII\ templates, and an ideal dataset for
construction of a new template would consist of complete UV and
optical spectra at high S/N, observed with a small spectroscopic
aperture to minimize starlight and narrow emission-line components.
New \HST\ observations are currently planned that will enable the
construction of a new \FeII\ template using quasi-simultaneous UV and
optical STIS data for the nearby Seyfert 1 galaxy Mrk 493 (program
GO-14744, PI:Park).

Calibration of \MgII-based SE BH mass estimators using available \HST\ archival spectra as well as our STIS data 
will be  presented in a future paper.
As an extension of the work of \citet{Wang+2009},
our STIS sample will provide an expanded BH mass range to calibrate
the \MgII\ virial relationship
and we will provide \linedisp\ and MAD-based calibrations as well as updated FWHM-based calibrations
using the  uniform measurement and analysis methods for spectral decompositions, uncertainty estimates, 
and Bayesian linear regression presented in this work.

\acknowledgments

We thank the anonymous referee for a prompt and constructive
report that has improved the paper.
Support for \HST\ program GO-12922 was provided by NASA through a grant 
from the Space Telescope Science Institute, which is operated by the 
Association of Universities for Research in Astronomy, Inc., under NASA 
contract NAS 5-26555.
DP acknowledges support through the EACOA Fellowship from The East Asian 
Core Observatories Association, which consists of National Astronomical 
Observatories, Chinese Academy of Science (NAOC), National Astronomical 
Observatory of Japan (NAOJ), Korean Astronomy and Space Science Institute 
(KASI), and Academia Sinica Institute of Astronomy and Astrophysics (ASIAA). 
Research by AJB is supported in part by NSF grant AST-1412693.
JHW acknowledges the support by the National Research Foundation of Korea 
grant funded by the Korea government (No. 2010-0027910 and No. 2016R1A2B3011457).
TT acknowledges support by the Packard Foundation through a Packard Research 
Fellowship and by the National Science Foundation through grant AST-1412315. 
VNB gratefully acknowledges assistance from a National Science Foundation (NSF) Research 
at Undergraduate Institutions (RUI) grant AST-1312296. Note that findings and conclusions 
do not necessarily represent views of the NSF. 
RJA was supported by FONDECYT grant number 1151408.
We thank Jonelle L. Walsh for help with the STIS data reduction.

\clearpage

\capstartfalse

\begin{deluxetable*}{lccccc}
	\tablecolumns{6}
	\tablewidth{0pt}
	\tablecaption{Optical spectral properties from H$\beta$ reverberation mapping} 
	\tablehead{ 
		\colhead{Object} &
		\colhead{$z$} &
		\colhead{$\tau_{\rm cent}$} & 
		\colhead{$\sigma_{\rm rms}$} &
		\colhead{$\log (M_{\rm BH}$/\msun)} & 
		\colhead{References} \\ 
		\colhead{} &
		\colhead{} &
		\colhead{(H$\beta$)} &
		\colhead{(H$\beta$)} &
		\colhead{(RM)} &
		\colhead{} \\
		\colhead{} &
		\colhead{} &
		\colhead{(days)} &
		\colhead{(\kms)} &
		\colhead{} &
		\colhead{} \\
		\colhead{(1)} &
		\colhead{(2)} &
		\colhead{(3)} &
		\colhead{(4)} &
		\colhead{(5)} &
		\colhead{(6)} 
	} 
	\startdata
	\multicolumn{6}{c}{Sample presented in P13\tablenotemark{a}}\\
	\\
	3C 120        & $0.03301$  &  $27.2^{+1.1}_{-1.1}$  &  $1514\pm65$  &  $ 7.80\pm 0.31$  &  6  \\
	3C 390.3      & $0.05610$  &  $23.60^{+6.45}_{-6.45}$  &  $3105\pm81$  &  $ 8.43\pm 0.33$  &  1  \\
	Ark 120       & $0.03230$  &  $39.05^{+4.57}_{-4.57}$  &  $1896\pm44$  &  $ 8.14\pm 0.32$  &  1  \\
	Fairall 9     & $0.04702$  &  $17.40^{+3.75}_{-3.75}$  &  $3787\pm197$  &  $ 8.38\pm 0.32$  &  1  \\
	Mrk 279       & $0.03045$  &  $16.70^{+3.90}_{-3.90}$  &  $1420\pm96$  &  $ 7.51\pm 0.33$  &  1  \\
	Mrk 290       & $0.02958$  &  $8.72^{+1.21}_{-1.02}$  &  $1609\pm47$  &  $ 7.36\pm 0.32$  &  4  \\
	Mrk 335       & $0.02578$  &  $14.1^{+0.4}_{-0.4}$  &  $1293\pm64$  &  $ 7.37\pm 0.31$  &  6  \\
	Mrk 509       & $0.03440$  &  $79.60^{+5.75}_{-5.75}$  &  $1276\pm28$  &  $ 8.12\pm 0.31$  &  1  \\
	Mrk 590       & $0.02638$  &  $24.23^{+2.11}_{-2.11}$  &  $1653\pm40$  &  $ 7.65\pm 0.32$  &  1  \\
	Mrk 817       & $0.03145$  &  $19.05^{+2.45}_{-2.45}$  &  $1636\pm57$  &  $ 7.66\pm 0.32$  &  1  \\
	NGC 3516      & $0.00884$  &  $11.68^{+1.02}_{-1.53}$  &  $1591\pm10$  &  $ 7.47\pm 0.31$  &  4  \\
	NGC 3783      & $0.00973$  &  $10.20^{+2.80}_{-2.80}$  &  $1753\pm141$  &  $ 7.44\pm 0.32$  &  1  \\
	NGC 4593      & $0.00900$  &  $3.73^{+0.75}_{-0.75}$  &  $1561\pm55$  &  $ 6.96\pm 0.32$  &  2  \\
	NGC 5548      & $0.01717$  &  $4.18^{+0.86}_{-1.30}$  &  $3900\pm266$  &  $ 7.80\pm 0.34$  &  3, 5  \\
	NGC 7469      & $0.01632$  &  $4.50^{+0.75}_{-0.75}$  &  $1456\pm207$  &  $ 7.05\pm 0.31$  &  1  \\
	PG 0026+129   & $0.14200$  &  $111.00^{+26.20}_{-26.20}$  &  $1773\pm285$  &  $ 8.56\pm 0.33$  &  1  \\
	PG 0052+251   & $0.15500$  &  $89.80^{+24.30}_{-24.30}$  &  $1783\pm86$  &  $ 8.54\pm 0.32$  &  1  \\
	PG 0804+761   & $0.10000$  &  $146.90^{+18.85}_{-18.85}$  &  $1971\pm105$  &  $ 8.81\pm 0.31$  &  1  \\
	PG 0953+414   & $0.23410$  &  $150.10^{+22.10}_{-22.10}$  &  $1306\pm144$  &  $ 8.41\pm 0.32$  &  1  \\
	PG 1226+023   & $0.15830$  &  $306.80^{+79.70}_{-79.70}$  &  $1777\pm150$  &  $ 8.92\pm 0.32$  &  1  \\
	PG 1229+204   & $0.06301$  &  $37.80^{+21.45}_{-21.45}$  &  $1385\pm111$  &  $ 7.83\pm 0.38$  &  1  \\
	PG 1307+085   & $0.15500$  &  $105.60^{+41.30}_{-41.30}$  &  $1820\pm122$  &  $ 8.61\pm 0.33$  &  1  \\
	PG 1426+015   & $0.08647$  &  $95.00^{+33.50}_{-33.50}$  &  $3442\pm308$  &  $ 9.08\pm 0.34$  &  1  \\
	PG 1613+658   & $0.12900$  &  $40.10^{+15.10}_{-15.10}$  &  $2547\pm342$  &  $ 8.42\pm 0.38$  &  1  \\
	PG 2130+099   & $0.06298$  &  $12.8^{+1.2}_{-0.9}$  &  $1825\pm65$  &  $ 7.63\pm 0.31$  &  6  \\
	\hline
	\multicolumn{6}{c}{New sample presented here}\\
	\\
	Arp 151       & $0.02109$  &  $3.99^{+0.49}_{-0.68}$  &  $1295\pm37$  &  $ 6.83\pm 0.32$  &  3, 5  \\
	Mrk 1310      & $0.01956$  &  $3.66^{+0.59}_{-0.61}$  &  $921\pm135$  &  $ 6.50\pm 0.34$  &  3, 5  \\
	Mrk 50        & $0.02343$  &  $10.64^{+0.82}_{-0.93}$  &  $1740\pm101$  &  $ 7.50\pm 0.32$  &  7  \\
	NGC 6814      & $0.00521$  &  $6.64^{+0.87}_{-0.90}$  &  $1697\pm224$  &  $ 7.28\pm 0.34$  &  3, 5  \\
	SBS 1116+583A & $0.02787$  &  $2.31^{+0.62}_{-0.49}$  &  $1550\pm310$  &  $ 6.74\pm 0.38$  &  3, 5  \\
	Zw 229-015    & $0.02788$  &  $3.86^{+0.69}_{-0.90}$  &  $1590\pm47$  &  $ 6.99\pm 0.32$  &  8  
	\enddata
	\label{tab:RMdata}
	\tablecomments{
		Col. (1) Name.
		Col. (2) Redshifts are from the NASA/IPAC Extragalactic Database (NED).
		Col. (3) Rest-frame H$\beta$ time lag measurements.
		Col. (4) Line dispersion (\linedisp) measured from rms spectra.
		Col. (5) \mbh\ estimates from reverberation mapping: 
		$M_{\rm BH}({\rm RM})=f{\rm VP_{BH}}=fc\tau_{\rm cent}\sigma_{\rm rms}^2/G$ 
		where the virial factor $f$ with its uncertainty is adopted from \citealt{Park+2012ApJS} and 
		\citealt{Woo+2010} (i.e., $\log f = 0.71 \pm 0.31$).
		Col. (6) References. 
		1. \citealt{Peterson+2004};
		2. \citealt{Denney+2006};
		3. \citealt{Bentz+2009};
		4. \citealt{Denney+2010};
		5. \citealt{Park+2012};
		6. \citealt{Grier+2012};
		7. \citealt{Barth+2011:mrk50};
		8. \citealt{Barth+2011:zw229}
	}
	\tablenotetext{a}{Note that the sample and measurements are from P13. 
		One difference at here is that the adopted uncertainty for the virial factor (i.e., $0.31$ dex) 
		has been added in quadrature to the final RM BH mass uncertainties, although this homoscedastic
	    uncertainty addition into dependent variables does not alter any of calibration results in this work, 
	    except for the values of intrinsic scatter term and slight changes of constrained uncertainty ranges of regression coefficients.}
\end{deluxetable*}

\begin{deluxetable*}{lccccc}
	\tablecolumns{6}
	\tablewidth{0pt}
	\tablecaption{Summary of \HST/STIS observations for the six new AGNs} 
	\tablehead{ 
		\colhead{Object} &
		\colhead{Observation date} &
		\colhead{Slit PA} & 
		\multicolumn{3}{c}{Total exposure time} \\ 
		\colhead{} &
		\colhead{} &
		\colhead{} &
		\colhead{G140L} &
		\colhead{G230L} &
		\colhead{G430L} \\
		\colhead{} &
		\colhead{} &
		\colhead{(deg)} &
		\colhead{(sec)} &
		\colhead{(sec)} &
		\colhead{(sec)} 
	} 
	\startdata
	Arp 151       & 2013-04-29  &  97.7    &  3801  &  2639  &  495  \\
	Mrk 1310      & 2013-06-07  &  70.8    &  2624  &  1255  &  360  \\
	Mrk 50        & 2012-12-12  &  -110.8  &  2624  &  1255  &  360  \\
	NGC 6814      & 2013-05-07  &  -149.6  &  2848  &  1299  &  540  \\
	SBS 1116+583A & 2013-07-12  &  28.9    &  3714  &  2648  &  600  \\
	Zw 229-015    & 2013-07-23  &  117.6   &  4302  &  2942  &  600  
	\enddata
	\label{tab:ObsDetailes}
\end{deluxetable*}

\begin{turnpage}
	\begin{deluxetable}{lclccccccccc}
		\tablecolumns{11}
		\tablewidth{0pt}
		\tablecaption{Ultraviolet spectral properties from \ion{C}{4} single-epoch estimates} 
		\tablehead{ 
			\colhead{Object} &
			\colhead{Telescope/Instrument} &
			\colhead{Date Observed} &
			\colhead{S/N} &
			\colhead{$E(B-V)$} &
			\colhead{$\log (\lambda L_{\lambda}/$\ergs$)$} & 
			\colhead{FWHM$_{\rm SE}$} &
			\colhead{$\sigma_{\rm SE}$} &
			\colhead{MAD$_{\rm SE}$} &
			\colhead{$\rho(W,X)$} &
			\colhead{$\rho(W,Y)$} & 
			\colhead{$\rho(W,Z)$} \\
			\colhead{} &
			\colhead{} &
			\colhead{} &
			\colhead{(1450 \AA~or 1700 \AA)} &
			\colhead{} &
			\colhead{(1350 \AA)} &
			\colhead{(\CIV)} &
			\colhead{(\CIV)} &
			\colhead{(\CIV)} &
			\colhead{} &
			\colhead{} &
			\colhead{} \\
			\colhead{} &
			\colhead{} &
			\colhead{} &
			\colhead{(pix$^{-1}$)} &
			\colhead{(mag)} &
			\colhead{} &
			\colhead{(\kms)} &
			\colhead{(\kms)} &
			\colhead{(\kms)} &
			\colhead{} &
			\colhead{} &
			\colhead{} \\
			\colhead{(1)} &
			\colhead{(2)} &
			\colhead{(3)} &
			\colhead{(4)} &
			\colhead{(5)} &
			\colhead{(6)} &
			\colhead{(7)} &
			\colhead{(8)} &
			\colhead{(9)} &
			\colhead{(10)} &
			\colhead{(11)} &
			\colhead{(12)} 
		} 
		\startdata
		\multicolumn{12}{c}{Sample presented in P13\tablenotemark{a}}\\
		\\
3C 120         &  \IUE/SWP   &  1994-02-19,27;1994-03-11 &  $ 12$  & 0.263 &  $44.399\pm0.021$  &  $3093\pm 291$  &  $3106\pm 157$  &  $2280\pm  74$  &  $-0.05$  &  $-0.35$  &  $-0.50$    \\
3C 390.3       &  \HST/FOS   &  1996-03-31               &  $ 18$  & 0.063 &  $43.869\pm0.003$  &  $5645\pm 202$  &  $6154\pm  65$  &  $4674\pm  42$  &  $-0.11$  &  $-0.30$  &  $-0.32$    \\
Ark 120        &  \HST/FOS   &  1995-07-29               &  $ 17$  & 0.114 &  $44.400\pm0.005$  &  $3471\pm 108$  &  $3219\pm  53$  &  $2345\pm  31$  &  $ 0.01$  &  $-0.30$  &  $-0.37$    \\
Fairall 9      &  \HST/FOS   &  1993-01-22               &  $ 24$  & 0.023 &  $44.442\pm0.004$  &  $2649\pm  77$  &  $2694\pm  20$  &  $1987\pm  13$  &  $ 0.04$  &  $-0.21$  &  $-0.17$    \\
Mrk 279        &  \HST/COS   &  2011-06-27               &  $  9$  & 0.014 &  $43.082\pm0.004$  &  $4093\pm 388$  &  $2973\pm  53$  &  $2202\pm  18$  &  $ 0.01$  &  $-0.04$  &  $-0.11$    \\
Mrk 290        &  \HST/COS   &  2009-10-28               &  $ 24$  & 0.014 &  $43.611\pm0.002$  &  $2052\pm  36$  &  $3531\pm  32$  &  $2544\pm  14$  &  $-0.00$  &  $-0.13$  &  $-0.20$    \\
Mrk 335        &  \HST/COS   &  2009-10-31;2010-02-08    &  $ 29$  & 0.032 &  $43.953\pm0.001$  &  $1772\pm  14$  &  $1876\pm  12$  &  $1311\pm   7$  &  $ 0.03$  &  $-0.02$  &  $ 0.02$    \\
Mrk 509        &  \HST/COS   &  2009-12-10,11            &  $107$  & 0.051 &  $44.675\pm0.001$  &  $3872\pm  18$  &  $3568\pm   9$  &  $2601\pm   6$  &  $ 0.02$  &  $-0.07$  &  $-0.03$    \\
Mrk 590        &  \IUE/SWP   &  1991-01-14               &  $ 17$  & 0.033 &  $44.094\pm0.007$  &  $5362\pm 266$  &  $3479\pm 165$  &  $2630\pm 139$  &  $-0.09$  &  $-0.12$  &  $-0.05$    \\
Mrk 817        &  \HST/COS   &  2009-08-04;2009-12-28    &  $ 38$  & 0.006 &  $44.326\pm0.001$  &  $4580\pm  48$  &  $3692\pm  23$  &  $2756\pm  14$  &  $-0.01$  &  $-0.25$  &  $-0.21$    \\
NGC 3516       &  \HST/COS   &  2010-10-04;2011-01-22    &  $ 20$  & 0.038 &  $42.615\pm0.002$  &  $2658\pm  34$  &  $4006\pm  49$  &  $2864\pm  29$  &  $-0.03$  &  $ 0.07$  &  $ 0.06$    \\
NGC 3783       &  \HST/COS   &  2011-05-26               &  $ 29$  & 0.105 &  $43.400\pm0.001$  &  $2656\pm 444$  &  $2774\pm  91$  &  $2014\pm   9$  &  $-0.01$  &  $-0.01$  &  $-0.17$    \\
NGC 4593       &  \HST/STIS  &  2002-06-23,24            &  $ 10$  & 0.022 &  $43.761\pm0.005$  &  $2952\pm 166$  &  $2946\pm 162$  &  $2135\pm  33$  &  $-0.01$  &  $-0.00$  &  $-0.02$    \\
NGC 5548       &  \HST/COS   &  2011-06-16,17            &  $ 36$  & 0.018 &  $43.822\pm0.001$  &  $1785\pm  82$  &  $4772\pm  80$  &  $3528\pm 102$  &  $ 0.02$  &  $-0.11$  &  $-0.02$    \\
NGC 7469       &  \HST/COS   &  2010-10-16               &  $ 32$  & 0.061 &  $43.909\pm0.001$  &  $2725\pm  66$  &  $2849\pm 237$  &  $2060\pm  15$  &  $-0.02$  &  $-0.01$  &  $-0.14$    \\
PG 0026+129    &  \HST/FOS   &  1994-11-27               &  $ 25$  & 0.063 &  $45.236\pm0.005$  &  $1604\pm  50$  &  $4965\pm 113$  &  $3610\pm  77$  &  $-0.04$  &  $-0.11$  &  $-0.22$    \\
PG 0052+251    &  \HST/FOS   &  1993-07-22               &  $ 21$  & 0.042 &  $45.292\pm0.004$  &  $5380\pm  87$  &  $4648\pm  50$  &  $3463\pm  30$  &  $-0.12$  &  $-0.54$  &  $-0.61$    \\
PG 0804+761    &  \HST/COS   &  2010-06-12               &  $ 34$  & 0.031 &  $45.493\pm0.001$  &  $3429\pm  23$  &  $2585\pm  20$  &  $1932\pm  13$  &  $ 0.04$  &  $ 0.14$  &  $ 0.17$    \\
PG 0953+414    &  \HST/FOS   &  1991-06-18               &  $ 18$  & 0.012 &  $45.629\pm0.005$  &  $3021\pm  74$  &  $3448\pm  55$  &  $2472\pm  35$  &  $-0.02$  &  $-0.43$  &  $-0.48$    \\
PG 1226+023    &  \HST/FOS   &  1991-01-14,15            &  $ 93$  & 0.018 &  $46.309\pm0.001$  &  $3609\pm  29$  &  $3513\pm  29$  &  $2595\pm  19$  &  $-0.18$  &  $-0.52$  &  $-0.60$    \\
PG 1229+204    &  \IUE/SWP   &  1982-05-01,02            &  $ 28$  & 0.024 &  $44.609\pm0.009$  &  $4023\pm 163$  &  $2621\pm  90$  &  $1989\pm  62$  &  $-0.29$  &  $-0.48$  &  $-0.49$    \\
PG 1307+085    &  \HST/FOS   &  1993-07-21               &  $ 14$  & 0.030 &  $45.113\pm0.006$  &  $3604\pm 111$  &  $4237\pm  80$  &  $3010\pm  54$  &  $-0.13$  &  $-0.57$  &  $-0.58$    \\
PG 1426+015    &  \IUE/SWP   &  1985-03-01,02            &  $ 45$  & 0.028 &  $45.263\pm0.004$  &  $4220\pm 258$  &  $4808\pm 305$  &  $3549\pm  95$  &  $-0.10$  &  $-0.23$  &  $-0.42$    \\
PG 1613+658    &  \HST/COS   &  2010-04-08,09,10         &  $ 37$  & 0.023 &  $45.488\pm0.001$  &  $6398\pm  51$  &  $4204\pm  17$  &  $3286\pm  13$  &  $-0.01$  &  $-0.10$  &  $-0.04$    \\
PG 2130+099    &  \HST/COS   &  2010-10-28               &  $ 22$  & 0.039 &  $44.447\pm0.001$  &  $2147\pm  18$  &  $2225\pm  47$  &  $1554\pm  21$  &  $ 0.02$  &  $-0.06$  &  $-0.07$    \\
		\hline
		\multicolumn{12}{c}{New sample presented here}\\
		\\
Arp 151        &  \HST/STIS  &  2013-04-29               &  $  6$  & 0.012 &  $41.791\pm0.017$  &  $1489\pm  26$  &  $2900\pm  61$  &  $1864\pm  35$  &  $-0.03$  &  $-0.38$  &  $-0.46$    \\
Mrk 1310       &  \HST/STIS  &  2013-06-07               &  $  5$  & 0.027 &  $41.715\pm0.025$  &  $1434\pm  78$  &  $2447\pm 108$  &  $1603\pm  54$  &  $ 0.00$  &  $-0.25$  &  $-0.31$    \\
Mrk 50         &  \HST/STIS  &  2012-12-12               &  $ 19$  & 0.015 &  $43.213\pm0.003$  &  $2807\pm  63$  &  $4443\pm 160$  &  $3140\pm 115$  &  $ 0.02$  &  $-0.13$  &  $-0.10$    \\
NGC 6814       &  \HST/STIS  &  2013-05-07               &  $  6$  & 0.164 &  $41.105\pm0.021$  &  $2651\pm 264$  &  $2804\pm 103$  &  $2096\pm  59$  &  $ 0.02$  &  $-0.07$  &  $-0.11$    \\
SBS 1116+583A       &  \HST/STIS  &  2013-07-12               &  $ 13$  & 0.010 &  $42.867\pm0.005$  &  $3253\pm 302$  &  $3315\pm 231$  &  $2302\pm 121$  &  $-0.04$  &  $-0.13$  &  $-0.15$    \\
Zw 229-015         &  \HST/STIS  &  2013-07-23               &  $ 17$  & 0.064 &  $43.129\pm0.007$  &  $2573\pm  71$  &  $2608\pm  56$  &  $1891\pm  33$  &  $-0.05$  &  $-0.20$  &  $-0.18$      
		\enddata
		\label{tab:UVmeasurement}
		\tablecomments{
			Col. (1) Name.
			Col. (2) Telescope/Instrument from which archival UV spectra were obtained. Note that the new COS spectra were obtained after 2009.
			Col. (3) Observation date for combined spectra.
			Col. (4) Signal-to-noise ratio per pixel at 1450 \AA~or 1700 \AA~in rest-frame.
			Col. (5) $E(B-V)$ are from the NASA/IPAC Extragalactic Database (NED) based on the recalibration of Schlafly \& Finkbeiner (2011).
			Col. (6) Continuum luminosity measured at 1350 \AA.
			Col. (7) FWHM measured from SE spectra.
			Col. (8) Line dispersion (\linedisp) measured from SE spectra.
			Col. (9) MAD (mean absolute deviation around weighted median) measured from SE spectra.
			Col. (10) correlation coefficient between measurement errors of $W$ and $X$ where $W=\log \lambda L_{\lambda}$ at 1350 \AA\ and $X=\log \rm FWHM_{\rm SE}$.
			Col. (11) correlation coefficient between measurement errors of $W$ and $Y$ where $Y=\log \sigma_{\rm SE}$.
			Col. (12) correlation coefficient between measurement errors of $W$ and $Z$ where $Z=\log \rm MAD_{\rm SE}$.
		}
	\tablenotetext{a}{Note that the sample and measurements are from P13. 
		One difference at here is that measurements for the MAD and error correlations 
		have been included.}
	\end{deluxetable}
\end{turnpage}

\begin{deluxetable*}{lccccccc}
	\tablecolumns{8}
	\tablewidth{0pc}
	\tablecaption{\CIV\ \mbh\ estimator calibration results\\
		$
		\log [M_{\rm BH} {\rm (RM)}/M_\odot] = ~ \alpha ~ + ~ \beta ~
		\log(L_{1350\textrm{\AA}}/10^{44}~\rm erg~s^{-1}) 
		~ + ~ \gamma ~ \log [\varDelta V(\textrm{\CIV})/1000~ \rm km~s^{-1}]
		$
	}
	\tablehead{
		\colhead{$\varDelta V(\textrm{\CIV})$}      & 
		\colhead{$\alpha$}    & 
		\colhead{$\beta$}     & 
		\colhead{$\gamma$}    & 
		\colhead{$\sigma_{{\mathop{\rm int}}}$} &
		\colhead{mean offset}    & 
		\colhead{1$\sigma$ scatter} &
		\colhead{Ref.} \\
		\colhead{} & 
		\colhead{} & 
		\colhead{} & 
		\colhead{} & 
		\colhead{} &
		\colhead{(dex)} & 
		\colhead{(dex)} &
		\colhead{} 
	}
	\startdata
	\multicolumn{7}{c}{Previous calibrations} \\
	\\
	$\sigma_{\rm line}$  & $6.73\pm0.01$ & $0.53$        & $2$           & $0.33$       & \nodata & \nodata & VP06\\
	FWHM                 & $6.66\pm0.01$ & $0.53$        & $2$           & $0.36$       & \nodata & \nodata & VP06\\
	$\sigma_{\rm line}$  & $6.71\pm0.07$ & $0.50\pm0.07$ & $2$           & $0.28\pm0.04$ & $0.00$  & $0.295$ & P13 \\
	FWHM                 & $7.48\pm0.24$ & $0.52\pm0.09$ & $0.56\pm0.48$ & $0.35\pm0.05$ & $0.00$  & $0.347$ & P13 \\
	\hline
	\hline
	\multicolumn{7}{c}{This work} \\
	\\
	$\sigma_{\rm line}$  & $6.90^{+0.35}_{-0.34}$ & $0.44^{+0.07}_{-0.07}$ & $1.66^{+0.65}_{-0.66}$ & $0.12^{+0.09}_{-0.06}$ & $0.01$  & $0.33$ &  \\
	FWHM                 & ${\bf 7.54}^{+0.26}_{-0.27}$ & ${\bf 0.45}^{+0.08}_{-0.08}$ & ${\bf 0.50}^{+0.55}_{-0.53}$ & $0.16^{+0.10}_{-0.08}$ & $0.00$  & $0.37$ &  best-fit\tablenotemark{a} \\
	MAD                  & $7.15^{+0.24}_{-0.25}$ & $0.42^{+0.07}_{-0.07}$ & $1.65^{+0.61}_{-0.62}$ & $0.12^{+0.09}_{-0.06}$ & $0.00$  & $0.33$ &  \\
	\hline
	\multicolumn{7}{c}{This work (fixing $\gamma = 2$)} \\
	\\
	$\sigma_{\rm line}$  & ${\bf 6.73}^{+0.07}_{-0.07}$ & ${\bf 0.43}^{+0.06}_{-0.06}$ & ${\bf 2}$ & $0.12^{+0.09}_{-0.06}$ & $0.01$  & $0.33$ & best-fit\tablenotemark{a} \\
	FWHM                 & $6.84^{+0.09}_{-0.09}$ & $0.33^{+0.07}_{-0.07}$ & $2$ & $0.22^{+0.11}_{-0.10}$ & $-0.01$  & $0.43$ &  \\
	MAD                  & ${\bf 7.01}^{+0.07}_{-0.07}$ & ${\bf 0.41}^{+0.06}_{-0.06}$ & ${\bf 2}$ & $0.12^{+0.09}_{-0.06}$ & $0.00$  & $0.33$ & best-fit\tablenotemark{a} \\
	\hline
	\multicolumn{7}{c}{This work (fixing $\beta = 0.5$)} \\
	\\
	$\sigma_{\rm line}$  & $6.99^{+0.34}_{-0.34}$ & $0.5$ & $1.49^{+0.63}_{-0.62}$ & $0.12^{+0.09}_{-0.06}$ & $0.01$  & $0.34$ &  \\
	FWHM                 & $7.62^{+0.23}_{-0.23}$ & $0.5$ & $0.31^{+0.46}_{-0.45}$ & $0.16^{+0.10}_{-0.08}$ & $0.01$  & $0.38$ &  \\
	MAD                  & $7.23^{+0.24}_{-0.24}$ & $0.5$ & $1.41^{+0.57}_{-0.58}$ & $0.12^{+0.09}_{-0.06}$ & $0.01$  & $0.34$ &  \\
	\hline
	\multicolumn{7}{c}{This work (fixing $\beta = 0.5$ and $\gamma = 2$)} \\
	\\
	$\sigma_{\rm line}$  & $6.72^{+0.07}_{-0.07}$ & $0.5$ & $2$ & $0.12^{+0.08}_{-0.06}$ & $0.01$  & $0.35$ &  \\
	FWHM                 & $6.82^{+0.09}_{-0.09}$ & $0.5$ & $2$ & $0.26^{+0.11}_{-0.11}$ & $0.00$  & $0.47$ &  \\
	MAD                  & $7.00^{+0.07}_{-0.07}$ & $0.5$ & $2$ & $0.12^{+0.09}_{-0.06}$ & $0.01$  & $0.35$ &  
	\enddata
	\label{tab:calibration}
	\tablecomments{The mean offset and $1\sigma$ scatter for our calibrations are measured from the average and standard deviation of 
		mass residuals between RM masses and calibrated SE masses, $\varDelta=\log M_{\rm BH} {\rm (RM)} - \log M_{\rm BH} {\rm (SE)}$.
		Note that the apparent big difference in $\sigma_{\rm int}$ estimates between the previous calibrations and this work
		is mostly due to the differences in the adopted RM mass error and statitcal model.
		The uncertainty of $\log f$ (i.e., 0.31 dex) is added in quadrature to the uncertainties of RM BH masses in this work.
		The standard deviation ($\sigma$) of the $t$ distribution is by definition different (larger) from that of Gaussian distribution due to the heavy-tail when the degrees-of-freedom parameter is small. In this case, the $\sigma_{\rm int}$ parameter of the $t$ distribution model is not the same as the data sperad ($\sigma$) of the $t$ distribution.
	}
	\tablenotetext{a}{We suggest these calibrations as the best \mbh\ estimators. 
     }
\end{deluxetable*}

\begin{deluxetable*}{lcccccc}
	\tablecolumns{7}
	\tablewidth{0pc}
	\tablecaption{Comparing calibration results with other linear regression methods\\
	}
	\tablehead{
		\colhead{Method}      & 
		\colhead{$\alpha$}    & 
		\colhead{$\beta$}     & 
		\colhead{$\gamma$}    & 
		\colhead{$\sigma_{{\mathop{\rm int}}}$} &
		\colhead{mean offset}    & 
		\colhead{1$\sigma$ scatter} \\
		\colhead{} & 
		\colhead{} & 
		\colhead{} & 
		\colhead{} & 
		\colhead{} &
		\colhead{(dex)} & 
		\colhead{(dex)} 
	}
	\startdata
	\texttt{Stan} (Bayesian)         & $7.54_{-0.27}^{+0.26}$ & $0.45_{-0.08}^{+0.08}$ & $0.50_{-0.53}^{+0.55}$ & $0.16_{-0.08}^{+0.10}$ & $0.00$  & $0.37$  \\
	\texttt{mlinmix\_err} (Bayesian) & $7.51_{-0.25}^{+0.25}$ & $0.43_{-0.07}^{+0.07}$ & $0.57_{-0.51}^{+0.50}$ & $0.24_{-0.09}^{+0.09}$ & $-0.00$  & $0.37$  \\
	\texttt{FITEXY} ($\chi^2$-based)       & $7.50\pm0.22$ & $0.43\pm0.08$ & $0.59\pm0.46$ & $0.20\pm0.10$ & $-0.00$  & $0.37$ 
	\enddata
	\label{tab:compare_methods}
	\tablecomments{For a consistent comparison, the exactly same methodology of \texttt{FITEXY} used by P13 is applied.}
\end{deluxetable*}

\capstarttrue

\end{CJK*}
\end{document}